\documentclass[twocolumn]{emulateapj}
\shorttitle{Bayesian Fitting of Dust SEDs}
\shortauthors{Kelly et al.}

\usepackage{url}

\def\mnras{MNRAS}
\def\apj{ApJ}
\def\aap{A\&A}
\def\apjl{ApJL}
\def\apjs{ApJS}

\def\pasp{PASP}
\def\araa{ARA\&A}
\def\qjras{QJRAS}

\def\icarus{Icarus}

\newcommand{\cmsq}{cm$^{-2}$}
\newcommand{\cmt}{cm$^{-3}$}

\newcommand{\N}{$N$}

\newcommand{\T}{$T$}
\newcommand{\bt}{$\beta$}
\newcommand{\Tbeta}{$T - \beta$}

\newcommand {\apgt} {\ {\raise-.5ex\hbox{$\buildrel>\over\sim$}}\ }
\newcommand {\aplt} {\ {\raise-.5ex\hbox{$\buildrel<\over\sim$}}\ } 

\newcommand {\chisq} {$\chi^2$}

\begin{document}

\title{Dust SEDs in the era of
  Herschel and Planck: a Hierarchical Bayesian fitting technique} 

\author{Brandon C. Kelly\altaffilmark{1,2}, Rahul Shetty\altaffilmark{3},
  Amelia M. Stutz\altaffilmark{4}, Jens Kauffmann\altaffilmark{5}, Alyssa
  A. Goodman\altaffilmark{1}, \and Ralf Launhardt\altaffilmark{4}}

\altaffiltext{1}{Harvard-Smithsonian Center for Astrophysics, 60
  Garden Street, Cambridge, MA 02138}
\altaffiltext{2}{Department of Physics, Broida Hall, University of California,
  Santa Barbara, CA 93106-9530}
\altaffiltext{3}{Zentrum f\"ur Astronomie der Universit\"at Heidelberg, Institut f\"ur Theoretische Astrophysik, Albert-Ueberle-Str. 2, 69120 Heidelberg, Germany}
\altaffiltext{4}{Max Planck Institut f\"ur Astronomie, K\"onigstuhl 17, 69117 Heidelberg, Germany}
\altaffiltext{5}{NASA JPL, 4800 Oak Grove Drive, Pasadena, CA 91109}





\begin{abstract}

We present a hierarchical Bayesian method for fitting infrared
spectral energy distributions (SEDs) of dust emission to observed
fluxes.  Under the standard assumption of optically thin single
temperature ($T$) sources the dust SED as represented by a power--law
modified black body is subject to a strong degeneracy between \T\ and
the spectral index \bt.  The traditional non-hierarchical approaches,
typically based on \chisq\ minimization, are severely limited by this
degeneracy, as it produces an artificial anti-correlation between \T\
and \bt\ even with modest levels of observational noise.  The
hierarchical Bayesian method rigorously and self-consistently treats
measurement uncertainties, including calibration and noise, resulting
in more precise SED fits.  As a result, the Bayesian fits do not
produce any spurious anti-correlations between the SED parameters due
to measurement uncertainty. We demonstrate that the Bayesian method is
substantially more accurate than the \chisq\ fit in recovering the SED
parameters, as well as the correlations between them.  As an
illustration, we apply our method to {\it Herschel} and submillimeter
ground-based observations of the star-forming Bok globule CB244.  This
source is a small, nearby molecular cloud containing a single low-mass
protostar and a starless core.  We find that \T\ and \bt\ are weakly
positively correlated -- in contradiction with the \chisq\ fits, which
indicate a \Tbeta\ anti-correlation from the same data-set.
Additionally, in comparison to the \chisq\ fits the Bayesian SED
parameter estimates exhibit a reduced range in values.
\end{abstract}


\section{Introduction}

\setcounter{footnote}{1}

\label{introsec}

Dust provides an important observational avenue for investigating a
variety of astrophysical environments.  Though dust amounts to only a
fraction ($\sim$1/100) of the gaseous mass, it is an efficient
radiator and thus provides clues to the energy content of the
interstellar medium (ISM).  Dust grains are heated by a variety of
sources, such as stellar radiation or collisions with gas, and
re-radiates most of this energy as thermal emission. For temperatures
characteristic of the ISM, thermal emission from dust is predominantly
at far infrared (IR) and submillimeter wavelengths.  Consequently,
properties of the far-IR and sub-mm spectral energy distribution (SED)
can provide vital information about the physical state of observed
systems.  For example, the amount of emergent IR flux scales
proportionally with the star formation rate in (ultra) luminous
infrared galaxies ([U]LIRGs, e.g., \citealt{Lagacheetal05}); SED
shapes indicate the evolutionary state of star forming cores within
molecular clouds \citep[e.g.,][]{Adamsetal87}; and grain evolution in
protoplanetary disks can be modeled using measured SEDS
\citep[e.g.,][]{Watsonetal07}.  Understanding dust SEDs therefore
represents a crucial step in numerous research fields spanning a wide
range of spatial scales.  Ground and space based telescopes, such as
{\it IRAS}, {\it ISO}, {\it Spitzer}, {\it Herschel}, {\it Planck},
SCUBA(2), MAMBO(2), SABOCA, and new instruments on {\it SOFIA} have
and continue to measure IR emission, with the aim of quantifying dust
SEDs.

At far-IR wavelengths ($\lambda$ \apgt\ 60 $\mu {\rm m}$) the shape of
SEDs due to thermal emission from dust is empirically found to be well
represented as a Planck function, $B_{\nu}(T)$, evaluated at the dust
temperature \T, modified by a power law in frequency, $\nu^{\beta}$
\citep{hild83}:
\begin{equation}
  S_{\nu} = \Omega N\kappa_0
  \left(\frac{\nu}{\nu_0}\right)^{\beta} B_{\nu}(T). \label{eq-modbbody}
\end{equation}
Here, $\Omega$ is the solid angle of the observing beam, $N$ is the
column density, $B_{\nu}(T)$ is the Planck function, and $\kappa_0
(\nu / \nu_0)^{\beta}$ is the opacity of the emitting dust. Note that
$\tau_0 = N \kappa_0$ is the optical depth at frequency $\nu_0$.  An
essential assumption in Equation 1 is that dust is optically thin, so
that $\tau(\nu) \ll 1$. Following convention, we assume that the
opacity $\kappa_0$ is known, and employ $\kappa_0 = 0.009 {\rm cm^2 /
g}$ \citep{ossenkopf1994}, which accounts for the dust-to-gas ratio,
so that the free parameter in the fit is the gas column density
$N$. The spectral index \bt\ determines the opacity $\kappa_{\nu}$ of
the dust, and encodes information about grain composition.  Usually,
\bt\ is found to be $\sim$2 for silicate and/or carbonaceous grain
composition common in the diffuse ISM \citep{Draine_and_Lee84}.

Numerous observational investigations have focused on measuring the
value of \bt.  Most observational analyses estimate \T\ and \bt\ by
employing a least-squares (\chisq) SED fit.  However, due to the
degeneracy between \T\ and \bt, known since the earliest efforts to
derive dust properties from a limited number of far-IR and
submillimeter measurements \citep{Keeneetal80}, \chisq\ fits
underestimating \bt\ will naturally overestimate
\T\ \citep{Blainetal03,Sajinaetal06}, or vice versa.  Erroneous
estimates of \T\ and/or \bt\ may arise due to the inaccurate
assumption of a constant temperature along the line-of-sight
\citep{Shettyetal09b}, or uncertainty in the flux measurements.
Additionally, the presence of internal sources may also yield an
artificial \Tbeta\ anti-correlation in the estimated SED
\citep{Malinenetal11}.  \citet{Shettyetal09a} demonstrate through
Monte Carlo simulations that modest noise levels can lead to spurious
\Tbeta\ anti-correlations.  Employing models of isothermal sources
with constant \bt, they show that due to uncertainties as low as 5\%,
\chisq\ fits produce erroneous \T\ and \bt\ estimates.  Moreover, the
fits result in an artificial \Tbeta\ anti-correlation which is
remarkably similar to the trend derived from observational
investigations.  \citet{Shettyetal09a} conclude that \chisq\ SED fits
cannot reliably reveal the true \Tbeta\ correlation under realistic
conditions where noise is present.

This \Tbeta\ degeneracy may have obscured the relationship between
dust temperature and composition present in astrophysical sources.
Observational investigations employing careful statistical analyses
also conclude that the \Tbeta\ anti-correlation derived from \chisq\
fits may be spurious.  \citet{Schneeetal10} demonstrate that though a
\chisq\ fit to fluxes from the starless core TMC-1C produces the
anti-correlation, due to the degeneracy between \T\ and \bt\ the data
is also consistent with a constant \bt\ throughout the source.  Using
radiative transfer modeling, \citet{Juvelaetal11} suggest that
line-of-sight temperature variations may be responsible for the
\chisq\ estimate of a decrease in \bt\ towards an internally heated
core.

Accurately measuring \bt\ and \T\ is essential for understanding grain
growth and evolution.  Under the dust coagulation scenario, a higher
frequency of dust agglomeration in dense regions results in increased
grain sizes.  In protoplanetary disks, \bt\ is usually measured to be
\aplt 1 \citep[e.g.][]{Miyake&Nakagawa93, Mannings&Emerson94,
  Draine06}, lower than the larger scale ISM value of \bt\ $\sim$2
\citep{Draine_and_Lee84}.  \citet{Goldsmithetal97} found some indications
of a decrease in \bt\ towards higher density gas in the Orion
molecular cloud.  In broad terms, temperatures are generally lower in
higher density regions (away from embedded sources) due to more
effective shielding from the ambient interstellar radiation field.
Thus, in the dust coagulation scenario the decrease in \bt\ towards
denser regions should be associated with a decrease in temperature and
an increase in density.

On the other hand, \chisq\ SED fitting from various sources -- from
starless cores to entire galaxies -- suggests that \bt\ increases with
decreasing \T.  Using {\it IRAS} and balloon-borne {\it PRONAOS}
observations of various sources, \citet{Dupacetal03} found an inverse
correlation between the spectral index and temperature, with \bt\
ranging from 0.8 to 2.4, and \T\ ranging from 11 to 80 K.
\citet{Dupacetal03} suggested that the hyperbolic shape of the \Tbeta\
anti-correlation may have a physical basis, such as a variation of
dust composition with temperature.  \citet{Yang&Phillips07} found a
similar trend from a sample of LIRGs, although \citet{hayward11} argue
that the optically thin isothermal SED model is inappropriate for
extra-galactic sources, at least for sub-mm galaxies.  More recently,
SEDs derived from {\it Herschel} and {\it Planck} observations have
also conveyed \Tbeta\ anti-correlations
\citep[e.g.][]{Andersonetal10,Paradis10,Planck2011coldcore}.  These
results are interpreted to be in agreement with laboratory studies:
laboratory measurements of some amorphous materials also show a
\Tbeta\ anti-correlation \citep[e.g.][]{Agladzeetal1996,boudet05}.
However, those laboratory results depend on grain composition and
wavelength, and do not consider how variations in (volume) densities
similar to the range found in the ISM can affect the results.  The
comparison is further complicated by the spurious anti-correlation
arising in \chisq\ fits due to noise \citep{Shettyetal09a}, and by
variations of temperature and dust composition along the line-of-sight
\citep{Shettyetal09b, Malinenetal11, Juvelaetal11}.

Properly treating the degeneracy between \T\ and \bt\ is thus a
necessity for accurately fitting dust SEDs and assessing any variation
of grain properties with temperature.  The motivation for this work
stems from the \Tbeta\ anti-correlation found in observational
analysis which appear to be very similar to the spurious
anti-correlation due solely to statistical uncertainties
\citep{Shettyetal09a}.  The main goal of this work is to develop a SED
fitting method that rigorously treats statistical uncertainties so
that spurious \Tbeta\ anti-correlations are avoided.  To satisfy these
conditions, we introduce a hierarchical Bayesian approach for fitting
an ensemble of dust SEDs.

Hierarchical\footnote{Also called ``multilevel'' modeling.} modeling
is a statistical framework that was developed to handle data analysis
problems with multiple stages
\citep[e.g.,][]{gelman2007}. Hierarchical models are preferred for
complex data analysis 
problems, as they are able to effectively handle multiple sources of
uncertainty at all stages of the data anaysis.  Sources of statistical
uncertainties include random noise and calibration (correlated)
errors, which can contribute both multiplicative and additive
components. Hierarchical models can account for correlated
uncertainties and degeneracies between parameters, and may thus avoid
any spurious correlations.  Moreover, Bayesian methods calculate the
probability distribution of the parameters given the measured data, so
that the uncertainties returned by the inference are rigorous and
well-defined.  This ensures that all sources of measurement error are
properly incorporated into the uncertainties in the estimated
parameters.  This is in stark contrast to more traditional frequentist
and approximate methods, such as the propagation of errors or the
treatment of the best-fit values as the true values.  Frequentist
approaches can lead to biases and incorrect quantification of
uncertainty in complex problems.  Bayesian methods have been shown to
avoid such biases for numerous problems in a variety of scientific
disciplines, and, as we demonstrate in this work, can 
effectively handle the degeneracy between \T\ and \bt\ in SED fitting.

This paper is organized as follows.  In the next section, we describe
our hierarchical Bayesian model, and contrast it with the \chisq\ fit.
In Section 3, we test the Bayesian method on model data, and compare
the results to a \chisq\ fit.  We apply the Bayesian fit to observed
fluxes from starforming core CB244 in Section 4.  After a brief
discussion on the interpretation of our results in Section 5, we
provide a summary in Section 6.

\section{The Hierarchical Statistical Model}\label{s-model}

In this section, we present a hierarchical Bayesian technique that
simultaneously estimates the values of the column density \N, spectral
index \bt, and temperature \T, as well as their joint distribution,
directly from a set of observed fluxes. Although hierarchical Bayesian
modeling is becoming more common in astrophysics
\citep[e.g.,][]{loredo2004,kelly2007,Hoggetal10,mandel2011}, it is not
nearly as widely employed in most astrophysical fields as traditional
single-level methods. In order to allow the reader to become familiar
with the hierarchical modeling approach, we begin by providing an
overview of the method in Section 2.1.  To guide readers familiar with
single-level frequentist methods, we also contrast it with the
minimized-\chisq\ fit, which is commonly employed for estimating \N,
\bt, and \T.  A thorough description of the particular hierarchical
Bayesian model we develop is subsequently given in Sections 2.2 and
2.3.  \citet{gelman04} provides a clear overview of Bayesian methods,
including hierarchical models, and \citet{carroll2006} is a good
reference on methods for dealing with measurement errors; both
references also discuss Markov Chain Monte Carlo (MCMC) methods.

The power of Bayesian analysis is reflected in its output, which for
our problem is a probability distribution for each value of $N,\beta,$
and $T$ for each pixel, as well as for the parameters defining the
joint distibution of $N,\beta,$ and $T$, given the measured
data.\footnote{Often in practice, Bayesian algorithms do not output
  the probability distribution of the parameters directly, but rather
  output a set of random draws of the parameters from their posterior
  probability distribution.}  The probability distribution of the
parameters given the observed data is called the {\it posterior}
distribution, and provides a complete and straight-forward description
of our uncertainty in the parameters. This is contrasted with the
output of the frequentist methods, such as $\chi^2$ or
maximum-likelihood estimates, as frequentist methods provide a point
estimate of the parameters and sometimes an estimate of a confidence
region. However, while Bayesian posteriors are exact, the frequentist analogues,
i.e., confidence regions, are often difficult to estimate for complex
problems, such as the one we are adressing.

\subsection{Basics of Hierarchical Modeling of Dust SEDs}

Traditionally, parameters for SED models have been estimated by
minimizing $\chi^2$. More recently, non-hierarchical Bayesian methods
have been used to estimate the parameters for individual pixels or
sources \citep{Paradis10}. These methods are appropriate if the goal
is to estimate the SED parameters for a single pixel or source, as
there is only one level to the data analysis problem: the measured
flux values are generated from the SED parameters for an individual
pixel or source. In this case, one only needs a measurement model for
the data, e.g., that the measured data are obtained by contaminating
the SED at the observational wavelengths with Gaussian measurement
noise. However, for most scientific problems of interest, the data
analysis problem has multiple stages, for which traditional
non-hierarchical methods, such as those based on $\chi^2$,
are not applicable and lead to biases. Within the context of the
problem that we are addressing, namely the distribution of parameters
for dust SEDs, there are at least two levels to the data analysis
problem. First, there is the level corresponding to how the SED
parameters for individual pixels or sources are generated from the
distribution of these parameters. Second, there is the level
corresponding to how the measured fluxes are generated from the SED
parameters for individual pixels or sources. The point of hierarchical
modeling is to model and fit both levels simultaneously.

Under the traditional non-hierarchical approach, statistical inference
at the first (distribution) level would be performed using the
best-fit results from the second (individual SED) level; i.e., the
distribution of the SED parameters is estimated directly from the
best-fit values for individual pixels or sources obtained by
minimizing $\chi^2$, effectively treating the best-fit values as if
they were the true values. However, these best-fit values are
estimated with error. The distribution of the estimates is the
convolution of the distribution of the true values of the SED
parameters with the error distribution of their estimates. Therefore,
the distribution of the quantities estimated using non-hierarchical
methods will always be a biased estimate of the distribution of the
true values, or rather the distribution of the values that would have
been obtained in the absence of measurement errors. Within the context
of the $\beta$--$T$ relationship, this implies that the distribution
of $\beta$ and $T$ estimated using the $\chi^2$-based estimates will
always be biased toward an anti-correlation, as the error distribution
of $\beta$ and $T$ is anti-correlated. This is a mathematical fact and
is true for any value of the $S/N$, although distributions inferred
from higher $S/N$ data will not be as biased. The reason for this is
because the traditional non-hierarchical methods do not effectively
treat errors at all levels of the data analysis problem.

Within the context of estimating SED parameters and their
distribution, a hierarchical model is constructed by invoking a model
for the distribution of SED parameters, as well as a model for the
measured data. The model for the distribution of SED parameters has
its own set of free parameters, and all that is assumed is the
functional form. For example, one could assume that the distribution
of SED parameters is a Gaussian distribution, and then the free
parameters would be the mean and covariance of the SED
parameters. Both the model for the distribution of the SED parameters,
and the SED parameters for individual pixels or sources, are fit
simultaneously. This approach is able to effectively handle
uncertainty at all levels of the data analysis problem, and thus does
not suffer from the biases that traditional non-hierarchical
approaches do when estimating the distribution of SED
parameters. Moreover, because one assumes a model for the distribution
of the SED parameters, and fits all of the pixels or sources
simultaneously, one is able to obtain more precise estimates of the
SED parameters for individual sources as all of the information is
pooled together.

The difference between the traditional and hierachical modeling
approaches may be more easily understood by using an analogy with
astronomical imaging. As an example, consider the $\beta$--$T$
distribution. The distribution of the estimated values of $\beta$ and
$T$ is the convolution of the distribution of the true values with
their banana-shaped error distribution. Therefore, the error distribution of $\beta$
and $T$ can be thought of as a `PSF' which acts on the image
of the distribution of the true values of the parameters in the
$\beta$--$T$ plane, causing the observed image of the parameters in
the $\beta$--$T$ plane to be a
blurred version of the true image. If all pixels or sources have the same value of
$\beta$ and $T$, then the image of the true values in the $\beta$--$T$
plane is that of a `pointsource', and the image of the estimated
values is just the banana-shaped `PSF' (i.e., the error distribution). If there is
any spread in the $\beta$ and $T$ values, then the true image in the
$\beta$--$T$ plane is that of an `extended source'. In order to
effectively estimate the image of the true distribution of the parameters in
the $\beta$--$T$ plane, it is necessary to deconvolve the image of the
observed distribution with the error distribution. As we will show in
\S~\ref{s-likfunc}, the methods which employ $\chi^2$-based estimates
implicitly assume that the distribution of the parameters in the
$\beta$--$T$ plane is uniform over all possible values. Similarly,
non-hierarchical Bayesian methods which
treat the pixels or sources independently also employ this
assumption. Because of the assumption of a uniform distribution, the
non-hierarchical methods are not able to deconvolve the image of the
estimates in the $\beta$--$T$ plane from the error distribution as
both the image of the true and estimated distribution are expected to
look the same under this assumption. Therefore, both the $\chi^2$ and non-hierarchical
Bayesian methods do not recover the true image of the distribution and
are biased. However, the hierarchical modeling framework is able to do the
deconvolution because it also employs a model for the `source' image,
i.e., the image of the true distribution in the $\beta$--$T$
plane. Therefore, the hierarchical model significantly improves on the
simpler methods, and provides more accurate estimates of the
$\beta$--$T$ distribution.

Another advantage of hierarchical modeling is that is it possible to
divide the measurement process into multiple levels. This therefore
enables us to treat both additive measurement noise and multiplicative
calibration errors. Traditional methods usually assume a simple
single-level measurement model, and dealing with multiple kinds of
measurement error can be difficult for these approaches. However, it
is easy to develop a hierarchical model which appropriately treats
multiple levels or sources of measurement error.

In this work, we develop a hierarchical model for the analysis of
far-IR SEDs of astronomical dust. We perform Bayesian inference on our
model, as the uncertainties at all levels of the model are exact and
straight-forward to interpret under the Bayesian approach. While it is
also possible to perform frequentist inference using a hierarchical
model, we do not do this as the uncertainties on the estimated
quantities are harder to estimate and interpret. Moreover,
hierarchical models lend themselves natually to MCMC algorithms, so
Bayesian methods are also computationally straight-forward.

\subsection{The Measurement Model}

  \label{s-measmod}

We model the dust SED as a modified black body, given by Equation
(\ref{eq-modbbody}). In addition, we multiply Equation
(\ref{eq-modbbody}) by a `color-correction' factor, which corrects the
measured flux values for the varying SED across the photometeric band,
and, if available, is usually given in a tabulated form in the
observer's manual for an instrument.  The color-correction factor is
given as a function of $\beta$ and $T$.

The SED fomulated by Equation (\ref{eq-modbbody}) is applicable to
optically thin dust emission along lines-of-sight with a single
temperature.  Consequently, sources with a significant amount of
high-density material where dust becomes optically thick, and/or with
large temperature gradients, will exhibit SEDs which cannot be
accurately described by Equation (\ref{eq-modbbody}). We will modify
our technique in future work to implement a more realistic model for
the SED in such cases; however, we note that the goal of this paper is
to develop a method that minimizes the effects of
statistical error on the scientific conclusion, and therefore isolates
the systematic errors, allowing for more direct investigatations into
their effects.  We address the issue of systematic errors from
inappropriate application of Equation \ref{eq-modbbody} further in
Section \ref{discussion}.

For $j = 1, \ldots, m$ observing bands, a map is measured for the
source having $i = 1, \ldots, n$ pixels. Denote the frequency of the
$j^{\rm th}$ band as $\nu_j$, and denote the measured flux density for
the $i^{\rm th}$ pixel observed at $\nu_j$ as $\hat{S}_{ij}$. Assuming
Equation (\ref{eq-modbbody}), the measured flux densities are assumed
to be related to the actual values according to the measurement
equation
  \begin{equation}
    \hat{S}_{ij} = \delta_j S_{\nu_j}(N_i,\beta_i,T_i) +
    \epsilon_{ij}. \label{eq-measeq}
  \end{equation}
Here, $\epsilon_{ij}$ is the random measurement error of the flux
density due to noise at frequency $\nu_{j}$ for the $i^{\rm th}$
pixel, and $\delta_j$ is the calibration error for the band
corresponding to $\nu_j$. Note that the calibration error is assumed
to be positive and the same for each pixel in the $j^{\rm th}$ band.

We assume that the noise is independent between pixels in the same map
and over different observing bands, and independent of the calibration
uncertainties. In addition, we assume that the calibration errors are
independent over the bands. Our assumption that the noise is
independent between bands and independent of the calibration error is
likely true. However, it is not true that the noise is independent
between pixels in the same band. This is because all images are
smoothed to the same resolution, therefore correlating the noise in
nearby pixels. While this can in theory be accounted for, it is
difficult to include in our statistical model and even more difficult
to develop an efficient MCMC sampler that accounts for
this. Therefore, for simplicity we ignore the correlations in the
noise among neighboring pixels. In addition, the calibration errors
are often correlated, and therefore our assumption that they are
independent may be incorrect. However, for simplicity, and because a
quantified summary of the correlations in $\delta_j$ is typically not
available, we assume the calibration uncertainties are independent.

We employ a robust statistical model for the measurement errors and
calibration uncertainties. This is to ensure that our conclusions are
not severely affected 
by our assumptions regarding these errors, as robust models
allow for outliers and other deviations from the modeling
assumptions. We model the distributions of $\epsilon_{ij}$ and
$\log \delta_j$, denoted as $p(\epsilon_{ij})$ and $p(\log \delta_j)$,
respectively, as having a Student's $t$-distribution with degrees of
freedom $d_{\rm meas}$ and $d_{\rm cal}$, zero mean, and scale parameters $\sigma_{ij}$ and
$\tau_j$. The distribution of the measurement noise is then
  \begin{eqnarray}
    \lefteqn{ p(\epsilon_{ij}) = } \nonumber \\
    & & \frac{\Gamma((d_{\rm meas} +1) /
      2)}{\Gamma(d_{\rm meas}/2)\sqrt{d_{\rm meas}\pi\sigma_{ij}^2}}
    \left(1 + \frac{\epsilon_{ij}^2}{\sigma^2_{ij}d_{\rm
          meas}}\right)^{-(d_{\rm meas}+1)/2} \label{eq-tdist1}
    \end{eqnarray}
    and the distribution of the calibration errors is
    \begin{eqnarray}
    \lefteqn{ p(\log \delta_{j}) =} \nonumber \\
    & & \frac{\Gamma((d_{\rm cal}+1) /
      2)}{\Gamma(d_{\rm cal}/2)\sqrt{d_{\rm cal}\pi\tau_j^2}} \left(1
      + \frac{(\log \delta_{j})^2}{\tau^2_{j}d_{\rm
          cal}}\right)^{-(d_{\rm cal}+1)/2} \label{eq-tdist2}. 
  \end{eqnarray}
Here, $\Gamma(\cdot)$ is the Gamma Function. In this work we assume
$d_{\rm meas} = 3$ and $d_{\rm cal} = 3$. The scale parameters,
$\sigma_{ij}$ and $\tau_j$, define the amplitude of the noise and calibration error 
distributions, respectively. In the limit that $d \rightarrow \infty$,
Equations (\ref{eq-tdist1}) and (\ref{eq-tdist2}) converge to a
Gaussian distribution with standard deviations $\sigma_{ij}$ and
$\tau_j$, respectively. However, unlike the Gaussian distribution, the
$t$-distribution has more probability in the tails of the
distribution, making it robust against outliers; indeed, a
$t$-distribution is commonly used when one requires a robust
statistical model \citep[e.g.,][]{gelman04}.

The $t$-distribution is often appropriate when the measurement errors
are assumed to be Gaussian, but their variance is only estimated and
therefore unknown. In this case, the $t$-distribution also
incorporates our uncertainty on the amplitude of the noise and
calibration errors. For example, modeling $\epsilon_{ij}$ as following
a Student's $t$-distribution corresponds to assuming that the noise is
Gaussian with standard deviation $\tilde{\sigma}_{ij}$, but we
estimate $\tilde{\sigma}_{ij}$ with $\sigma_{ij}$. This is an
appropriate model as the values of $\sigma_{ij}$ are typically
estimated from the maps themselves, e.g., by taking the dispersion in
the pixel values in a region of the image relatively free of
emission. The amplitude of the calibration uncertainties, 
$\tau_j$, are usually available from an observer's manual, but still
have some uncertainty associated with them as they are often estimated
by comparing a model fit to a calibration source.

When we assume that the unknown $\tilde{\sigma}^2_{ij}$ follows a
Scaled Inverse-$\chi^2$ distribution\footnote{The Inverse-$\chi^2$
  distribution is very similar to a $\chi^2$ distribution and is
  commonly used to represent uncertainty in a variance parameter.}
with $d_{\rm meas}$ degrees of freedom and scale parameter
$\sigma^2_{ij}$, then $\epsilon_{ij}$ marginally follows a Student's
$t$-distribution with $d_{\rm meas}$ degrees of freedom and scale
parameter $\sigma^2_{ij}$. In other words, Equation (\ref{eq-tdist1})
is equivalent to the following:
\begin{eqnarray}
  \epsilon_{ij}|\tilde{\sigma}^2_{ij} & \sim & N(0,\tilde{\sigma}^2_{ij}) \label{eq-hmeas1} \\
  \tilde{\sigma}^2_{ij} & \sim & \mbox{\rm Inv-}\chi^2(d_{\rm meas},\sigma^2_{ij}) \label{eq-hmeas2}.
\end{eqnarray}
Here, the notation $x|y \sim p(x|y)$ means that given $y$, $x$ is drawn from the
conditional probability distribution of $x$ given $y$, $p(x|y)$. In
addition, $N(\mu,V)$ is a Gaussian distribution with mean 
$\mu$ and variance $V$, and Inv-$\chi^2(f,s^2)$ is a scaled
Inverse-$\chi^2$ distribution with $f$ degrees of freedom and scale
parameter $s$. 

\begin{figure*}
  \includegraphics[scale=0.33,angle=90]{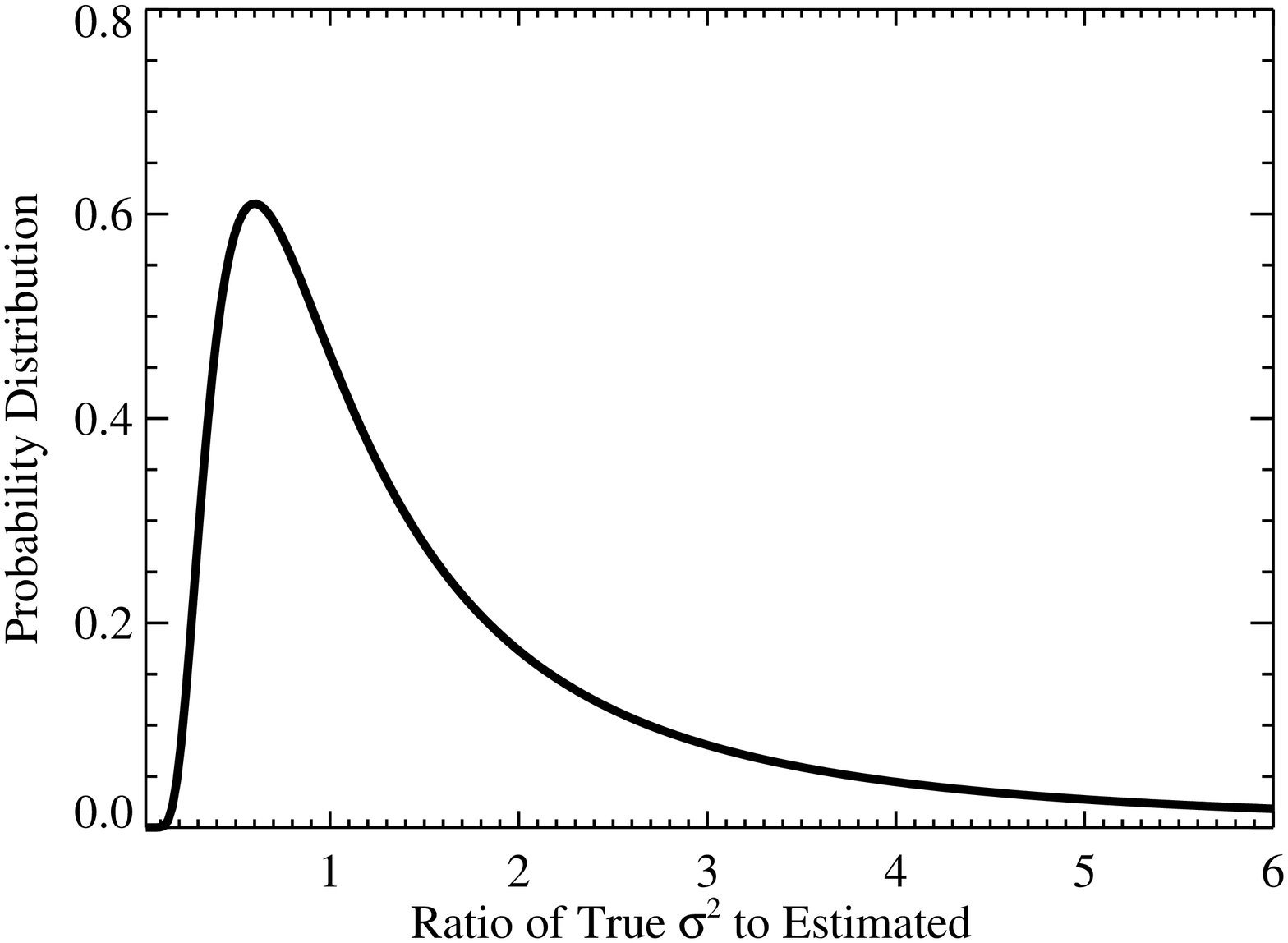}
  \includegraphics[scale=0.33,angle=90]{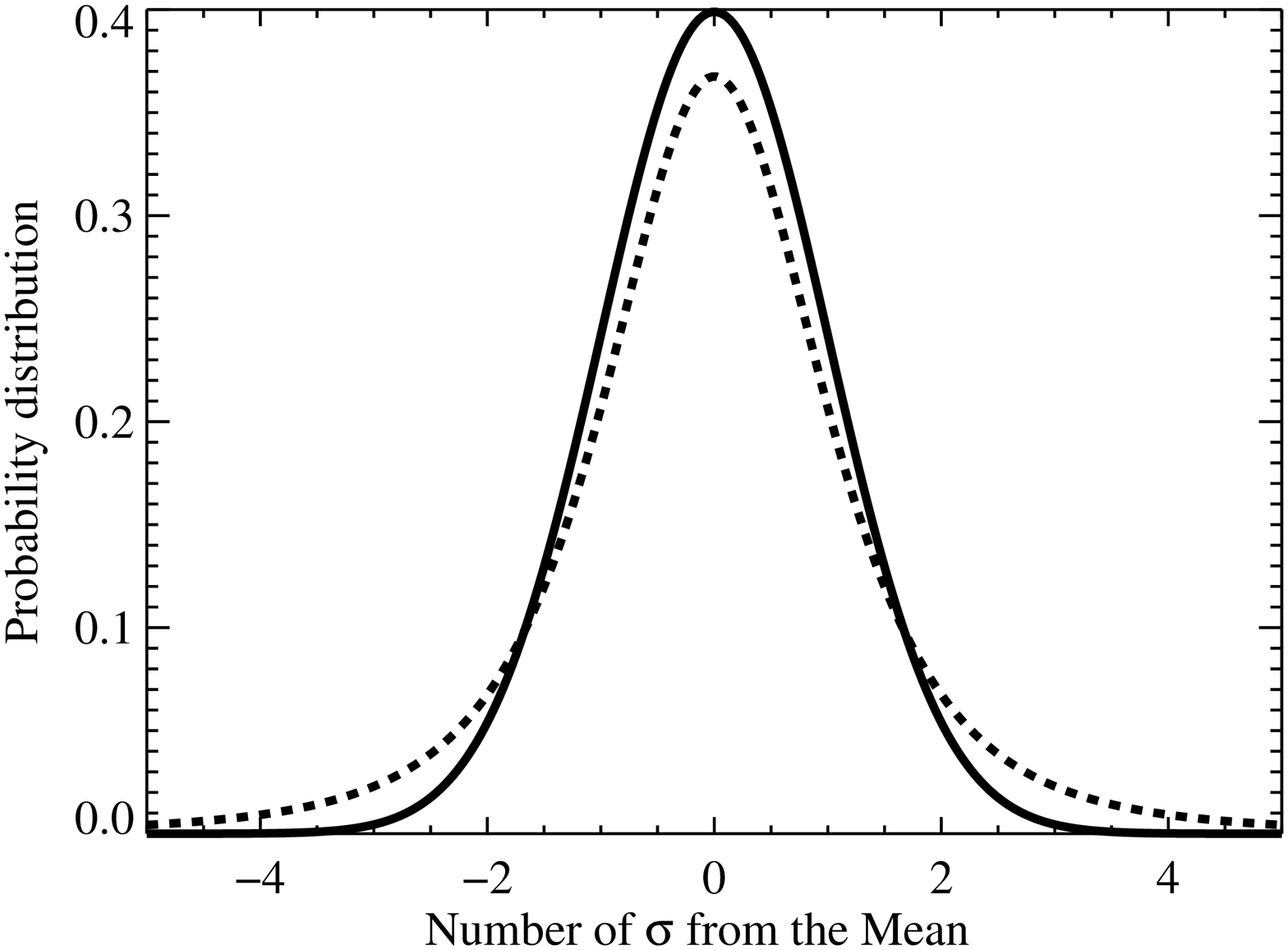}
  \caption{The scaled Inverse-$\chi^2$ distribution with $d = 3$
    degrees of freedom (left), and the Student's $t$-distribution with
    $d = 3$ degrees of freedom (right, dashed line) compared with a
    Gaussian distribution (solid line). The Student's $t$-distribution
    is appropriate when the errors are assumed to be Gaussian, but the
    uncertainty on the variances of the errors can be modeled as a
    scaled Inverse-$\chi^2$ distribution. The additional uncertainty
    on the amplitude of the errors is reflected in the thicker tails
    of the $t$-distribution, making the $t$-model more robust against
    outliers. In this work we model both the measurement noise and
    calibration uncertainties as following a Student's
    $t$-distribution with $d_{\rm meas} = 3$ and $d_{\rm cal} = 3$
    degrees of freedom for the measurement noise and calibration
    uncertainties, respectively, as the amplitudes of both the the
    noise and calibration errors are only estimated; the $t$-model
    therefore makes our method robust against outliers.}
  \label{f-inversechi2}
\end{figure*}

In Figure \ref{f-inversechi2} we show the scaled
Inverse-$\chi^2$ distribution with 3 degrees of freedom. In
addition, we compare the $t$-distribution with 3 degrees of freedom to a Gaussian
distribution. We chose a value of $d_{\rm meas} = d_{\rm cal} = 3$ for our measurement error model because 
it is the smallest value for the degrees of freedom that ensures that
the Student's 
$t$-distribution has a finite mean and variance. A value of 3 degrees
of freedom implicitly assumes that $\sim 6\%$ of the data will be outliers by
more than $3\sigma$. Alternatively, a value of 3 degrees of freedom can be
interpreted as assuming a factor
of $\sim 2$ uncertainty on the amplitude of the noise. While this may
be somewhat excessive, we prefer this 
conservative approach to make our results robust against inaccuracies
in our statistical model, as well as unknown systematic effects. This
robustness is illustrated in Figure \ref{f-inversechi2}, which shows
that outliers are expected in the $t$ measurement error model as
opposed to the Gaussian model, and the outliers are therefore down-weighted to ensure
that they do not have an excessive influence on the results. The
results are not strongly affected by changes in the degrees of
freedom which are less than an order of magnitude, and values of $d <
10$ are typical for robust models \citep[e.g.,][]{gelman04}.

\subsection{The Distribution Model and Posterior Distribution}

\label{s-likfunc}

In this section we derive the probability distribution for the
individual values of $N_i,T_i,\beta_i$, and the parameters for their
joint distribution, given a set of measured dust maps, assuming the
measurement model in \S~\ref{s-measmod}. This is called the `posterior
distribution', and is the basis for our Bayesian approach. In order to
simultaneously estimate the values of $N,T,$ and $\beta$ for each
pixel, as well as their joint distribution, we introduce the
additional step of assuming a parameteric form for the joint
distribution for $N,T,$ and $\beta$. Denote the set of
parameters\footnote{The notation $\theta$ is used as a short-hand way
  of denoting the set of parameters for the distribution of $N,T,$ and
  $\beta$. Once we choose a particular distribution, then $\theta$
  contains the parameters for this distribution.  At this point the
  derivation is still general so $\theta$ is left unspecified. Later
  we will model the distribution of $\log N, \log T,$ and $\beta$ as a
  Student's $t$-distribution, so $\theta = (\mu,\Sigma)$, where $\mu$
  is the mean vector and $\Sigma$ is proportional to the covariance
  matrix.} for the distribution of $N,T,$ and $\beta$ for the source
as $\theta$, and denote the distribution of these quantities as
$p(N,T,\beta|\theta)$. Then, the posterior distribution is
\begin{equation}
p({\bf N, T, \beta}, \theta|\hat{\bf S}) \propto p(\theta) p({\bf
  N,T,\beta}|\theta) p(\hat{\bf S}|{\bf N,T,\beta}), \label{eq-post1} 
\end{equation}
where ${\bf N,T}$, and ${\bf \beta}$ are vectors containing the values
of column density, temperature, and spectral index for the $n$ pixels,
$\hat{\bf S}$ is an $n \times m$ matrix containing the measured flux
densities, and $p(\theta)$ is the prior distribution on $\theta$. The
term $p(\hat{\bf S}|{\bf N,T,\beta})$ is the likelihood function of
the measured data. The quantity $p({\bf N,T,\beta}|\theta)$ defines
the distribution of $N,T,$ and $\beta$, while the likelihood function
$p(\hat{\bf S}|{\bf N,T,\beta})$ is defined by the measurement model
for the maps, i.e., how $N$, $T$ and $\beta$ generate the measured
flux densities for each map.

The existence of the calibration uncertainties makes calculation of
Equation (\ref{eq-post1}) difficult, as there is no analytical form
for the likelihood function of the measured data, $p(\hat{\bf S}|{\bf
  N}, {\bf T},{\bf \beta})$. However, it is 
straightforward to calculate the likelihood function for the case of
fixed calibration uncertainty, $p(\hat{\bf S}|{\bf N}, {\bf T},{\bf
  \beta}, {\bf \delta})$. For our $t$ model for the measurement noise,
the likelihood function at fixed $\delta$ is 
\begin{eqnarray}
 \lefteqn{ p(\hat{S}_{ij}|N_i,T_i,\beta_i,\delta_j) \propto} \nonumber \\
 & & \frac{1}{\sigma_{ij}} \left[1 + \frac{(\hat{S}_{ij} - \delta_j
     S_{\nu_j}(N_i,\beta_i,T_i))^2}{\sigma^2_{ij}d_{\rm meas}}\right]^{-(d_{\rm meas}+1)/2}. \label{eq-tdist_shat}  
\end{eqnarray}
The actual measured data likelihood
function, $p(\hat{\bf S}|{\bf N}, {\bf T},{\bf \beta})$, is then
obtained by averaging $p(\hat{\bf S}|{\bf N}, {\bf T},{\bf
  \beta}, {\bf \delta})$ over the distribution of ${\bf
  \delta}$. Therefore, in order to derive Equation (\ref{eq-post1}),
we first start with the posterior distribution that we would have
obtained if we treated the
calibration uncertainties as additional parameters:
\begin{eqnarray}
\lefteqn{p({\bf N, T, \beta}, \theta, \log \delta|\hat{\bf S})
  \propto} \nonumber \\
 & & p(\theta) \left[ \prod_{i=1}^n p(N_i,T_i,\beta_i|\theta) \right]
  \prod_{j=1}^{m} \left[ p(\log \delta_j)
  \prod_{i=1}^n p(\hat{S}_{ij}|N_i,T_i,\beta_i,\delta_j) \right]. \label{eq-post2} 
\end{eqnarray}
Then, Equation (\ref{eq-post1}) is obtained by integrating Equation
(\ref{eq-post2}) over $\log \delta_j$ for $j = 1, \ldots, m$. 

In this work we model the distribution of $\log N$, $\log T$, and
$\beta$ as a multivariate Student's t-distribution with $d = 8$
degrees of freedom:
\begin{eqnarray}
\lefteqn{ p(\log N_i, \log T_i, \beta_i|\mu,\Sigma) \propto} \nonumber \\
& & \frac{1}{|\Sigma|^{1/2}} \left[ 1 + \frac{1}{d}({\bf x}_i - \mu)^T
  \Sigma^{-1} ({\bf x}_i - \mu) \right]^{-(d+3) /
  2} \label{eq-tdist_pop} \\ 
{\bf x_i} & = & (\log N_i, \log T_i, \beta_i) \label{eq-xdefine}.
\end{eqnarray}
Here, ${\bf x}^T$ denotes the transpose of ${\bf x}$, $\theta = (\mu,
\Sigma)$, $\mu$ is the model mean value of $(\log N, \log T, \beta)$,
and $\Sigma$ is proportional to the 
model covariance matrix of $(\log N, \log T, \beta)$. Our reasons for
using $d = 8$ are similar to the case of
modeling the measurement errors; we want to use a distribution that is
robust against outlying values of $N,T,$ or $\beta$. A value of $d =
8$ implies that we expect about $\sim 1.6\%$ of the data to be
outliers by more than $3\sigma$. We consider this to be a reasonable
choice, but as with the measurement model, our results are not
strongly affected by the choice of $d$. Under the $t$-model,
 one can now use Equations
(\ref{eq-tdist_shat})--(\ref{eq-xdefine}) in combination with Equation
(\ref{eq-tdist2}) and a prior distribution $p(\theta)$ to calculate
the posterior distribution. 

In this work we assume a uniform prior on $\mu$. For the prior on
$\Sigma$, we use the so-called `seperation strategy' prior developed
by \citet{sepstrag}. This prior is based on the decomposition
\begin{equation} 
  \Sigma = S R S,
  \label{eq-sepstrag}
\end{equation} 
where $S$ is the diagonal matrix of standard deviations
and $R$ is the correlation matrix. The seperation strategy works by
placing independent priors on the standard deviations and
correlations. In this work we place a normal prior on the elements of
$\log S$ centered at the values inferred from the $\chi^2$ estimates
with variance equal to 100; this is an extremely broad prior giving
nearly equal weight to most reasonable values of the dispersion in
$\log N, \beta,$ and $\log T$. Following \citet{sepstrag}, we place an
inverse-Wishart prior on $R$ with four degrees of freedom. Under this
choice of prior, the marginal prior distributions for the correlations
between $\log N, \beta,$ and $\log T$ are uniform over $[-1,1]$,
reflecting our prior assumption that all values of the correlations
are equally likely.

The traditional non-hierarchical methods can be derived as a
special case under our hierarchical Bayesian model. When
the errors are assumed to be Gaussian ($d \rightarrow \infty$),
one ignores calibration uncertainties ($\delta_j = 1$), and when
one assumes that $\log N, \log T,$ and $\beta$ are independently and uniformly
distributed over all possible values ($\Sigma \rightarrow
\infty$), the posterior distribution becomes 
\begin{equation}
  p({\bf N}, {\bf T}, {\bf \beta}|\hat{\bf S}) \propto \prod_{i=1}^n
  \exp(-\chi_i^2 / 2), \label{eq-chi2_post} 
\end{equation}
where
\begin{equation}
  \chi_i^2 = \sum_{j=1}^m \left[\frac{\hat{S}_{ij} -
      S_{\nu_j}(N_i,\beta_i,T_i)}{\sigma_{ij}}\right]^2. \label{eq-chi2} 
\end{equation}

If the estimated values of $N,T,$ and $\beta$ are constrained to fall
within some range, say $0 < \beta < 5$, then Equation
(\ref{eq-chi2_post}) is derived from the assumption of a uniform
distribution over this range. Under Equation (\ref{eq-chi2_post}) the
values of $N_i,T_i,$ and $\beta_i$ are independent in their posterior
probability distribution, and thus their best-fit values can be
obtained independently for each pixel. The $\chi^2$-based estimators
are those that maximize the posterior distribution under the
simplifying assumptions that the errors are Gaussian, that there are
no calibration uncertainties, and that $\log N, \log T,$ and $\beta$
are independently and uniformly distributed over all possible
values. Similarly, non-hierarchical Bayesian methods based on Equation
(\ref{eq-chi2_post}) also make these same assumptions
\citep{Paradis10}. However, it is not true that all three of these
assumptions hold. In particular, the assumption that $\log N, \log T,$
and $\beta$ are independently and uniformly distributed over some
range of values is a very strong assumption and leads to estimates of
these quantities that are independent for each pixel, and thus the
estimated quantities are overdispersed compared to their true
values. This overdispersion biases the inferred correlation between
$\beta$ and $T$ toward an anti-correlation, such as that described by
\citet{Shettyetal09a}. Indeed, the very fact that an anti-correlation
between $\beta$ and $T$ is inferred from such methods shows that their
assumption of statistical independence is violated, as statistically
independent quantities must be uncorrelated. Therefore, such methods
are not self-consistent. Our hierarchical Bayesian approach
significantly ameliorates these problems by using a more realistic
model for the distribution of $\log N, \log T,$ and $\beta$, for the
measurement errors, and by including the calibration uncertainties.

The posterior distribution, given by Equation (\ref{eq-post1}),
completely summarizes our information on ${\bf N}, {\bf T}, {\bf
  \beta}, \mu,$ and $\Sigma$. We can then use Equation
(\ref{eq-post1}) to compute estimates of these parameters, and
summarize our uncertainty on $\mu$ and $\Sigma$. However, there are a
large number of parameters involved in Equation (\ref{eq-post1}), as
each source is assumed to have its own value of $N$, $T$, and
$\beta$. Therefore, we will not be able to compute Equation
(\ref{eq-post1}) on a grid, as there are $3n + 9$ free
parameters. Because of this, we employ MCMC methods to obtain a set of
random draws of ${\bf N}, {\bf T}, {\bf \beta}, \mu,$ and $\Sigma$
from the posterior distribution. In addition, it is difficult, if not
impossible, to analytically integrate Equation (\ref{eq-post2}) over
each $\delta_j$, which is necessary in order to compute Equation
(\ref{eq-post1}). For large values of $n$, it is computationally
intensive to do the integration over $\log \delta_j$ numerically
because of the product over the $n$ data points. In order to make the
computation of the posterior distribution tractable, we instead
consider the unknown values of the calibration errors, $\delta_j$, to
be additional parameters and work directly with Equation
(\ref{eq-post2}) instead of Equation (\ref{eq-post1}). Under this
strategy, the values of $\log \delta_j$ are additional parameters that
are also estimated at each stage of our MCMC sampler. However, because
the values of the calibration errors are of no scientific interest
(i.e., $\delta_j$ is a `nuisance' parameter), we simply discard the
values of $\log \delta_j$ returned by our MCMC sampler, thus
marginalizing over them. The result is a set of random draws of ${\bf
  N}, {\bf T}, {\bf \beta},$ and $\theta$ from Equation
(\ref{eq-post1}).

We construct our MCMC sampler using a combination of
Metropolis-Hastings updates and Gibbs sampling, where the Gibbs
updates are used whenever possible. In addition, to remove
degeneracies in some of the parameters, and thus to increase the
efficiency of our MCMC sampler, we employ an Ancillarity-Sufficiency
Interweaving Strategy \citep{asis} with respect to the calibration
uncertainties.  Implementing the interweaving strategy is necessary as
we could not get our MCMC sampler to converge without implementing
it. Further technical details of our MCMC sampler are given in
\citet{kelly_discuss}. All statistical inference is then done using
the random samples of ${\bf N}, {\bf T}, {\bf \beta}, \mu,$ and
$\Sigma$ generated from our MCMC algorithm.

Before concluding this section, we wish to make a comment on the
sensitivity of the Bayesian fits to the assumed population model. 
The true distribution for $\log N,\log T,$ and $\beta$ is unlikely to
follow a $t$-distribution (or Gaussian, for that matter), but it is
unlikely that errors due to this mismatch will have a significant
effect on our results. This is because we are primarily interested in
the moments of the data (e.g., the correlation between $T$ and
$\beta$), and simple models such as the $t$-distribution often enable
us to adequately recover them. Moreover, most of our analysis relies
on analyzing the values of $N_i,T_i,$ and $\beta_i$ returned by our
MCMC sampler, which have their values `corrected' relative to the
$\chi^2$-based estimates under the assumption that they come from a
common distribution, which we assume can be approximated as a
$t$-distribution. The degree to which the values of $N, T,$ and $\beta$ are corrected for any given pixel
depends on the $S/N$ of that pixel. When the $S/N$ is high, the
correction is small as the information on the values of $N_i,T_i,$ and
$\beta_i$ is dominated by the information from the data. In this
case, the estimated values of $N_i, T_i,$ and $\beta_i$ for that pixel
are insensitive to the choice of model for the joint distribution and will be similar to those obtained by minimizing $\chi^2$. By
focusing our analysis on the values of ${\bf N},{\bf T},$ and ${\bf
  \beta}$ that are returned by our Bayesian approach, instead of the
values of $\mu$ and $\Sigma$, we are less sensitive to differences
between our model and the true distribution. In general, so long as
the distribution of $N,T,$ and $\beta$ does not exhibit multiple modes
seperated by large distances, the scientific conclusions should not be
very sensitive to error due to mispecifying the statistical model. We
further explore this issue in the following section.

\section{Tests of hierarchical Bayesian method on simulated data}\label{testsec}

In order to illustrate the effectiveness of our Bayesian approach, and
its improvement over traditional \chisq-based techniques, we apply
our method to two simulated data sets. The first simulated data set is
a source with mean $T \sim 20$ K, and the second is a source with mean $T \sim
80$ K. We simulate values of the quantity $C =
N \Omega \kappa_0$ from a mixture of two log-normal distributions:
\begin{equation}
  p(\log C) = \pi \phi(\log C|\bar{C}_1,v^2_1) + (1 - \pi) \phi(\log
  C|\bar{C}_2,v^2_2). \label{eq-mixnorm}
\end{equation}
Here, $\phi(\log C|\bar{C},v^2)$ denotes a Gaussian distribution with mean
$\bar{C}$ and variance $v^2$ as a function of $\log C$. For both the cooler and warmer source we set $\pi = 0.4,
\bar{C}_1 = 7.6, \bar{C}_2 = 8.6, v_1 = 0.4,$ and $v_2 = 0.15$. These
values were chosen to give column densities similar to that observed for CB
244 as observed by {\it Herschel}, which we analyze in
\S~\ref{s-obssec}. We
simulate values of temperature from a mixture of two Gamma
distributions:
\begin{equation}
  p(T) = \pi \Gamma(T|k_1,T_1^*) + (1 - \pi)
  \Gamma(T|k_2,T^*_2). \label{eq-mixgam}
\end{equation}
Here, $\Gamma(T|k,T^*)$ is a Gamma distribution with shape parameter
$k$ and scale parameter $T^*$ as a function of $T$; note that the mean
and standard deviation of the Gamma distribution is $T^*k$ and
$T^*\sqrt{k}$, respectively. For the cooler simulated source we set
$k_1 = 500, T^*_1 = 0.03, k_2 = 100,$ and $T^*_2 = 0.2$. For the
hotter simulated source we set $k_1 = 500, T^*_1 = 0.15, k_2 = 1000,$
and $T^*_2 = 0.08$. Finally, for both sources we simulated values of
$\beta$ at fixed temperature from a Gaussian distribution with mean $A
+ B \log T$ and standard deviation 0.1. We set $B = 0.5$, and $A$ is
chosen to give a mean value of $\langle \beta \rangle= 2$. In our
simulations $\beta$ and $T$ are weakly and positively correlated by
construction.

We note that our model of a Student's $t$-distribution for the joint
distribution of $\log N, \log T,$ and $\beta$ is violated in these
simulations.  In fact, the distribution of $\log N$ is bimodal, as is
the distribution of $\log T$ for the cooler simulated
source.  These simulations will provide a good test of the
sensitivity of our results to our assumption of a Student's
$t$-distribution for $(\log N,\log T,\beta)$.

We simulated flux values for $n = 1000$ data points at observational
wavelengths $\lambda = 100, 160, 250, 350,$ and
$500$ \micron, corresponding to those employed by the {\it PACS} and {\it
  SPIRE} instruments onboard {\it Herschel}. We multiplied the flux
points in each band by a constant calibration error; the calibration
errors were the same for every pixel in a given band, but differed over
the five bands. The calibration errors were drawn from a log-normal
distribution with standard deviations of $2.75\%, 4.15\%, 7\%, 7\%,$
and $7\%$ for each band, respectively. These values were chosen to be
equal to the official calibration uncertainties for {\it PACS} and
{\it SPIRE} for a point source. 
To all fluxes, we also added noise drawn from Gaussian distributions with
standard deviations $\sigma_{j} = [2.2,
  3.3, 5.2, 3.7, 2.2] \times 10^{-5}\ {\rm Jy\ arcsec^{-2}}$ for each of the five bands,
respectively. These values were chosen to be similar to those observed
in the {\it Herschel} observation of CB244 \citep{Stutzetal10}, which
we apply our method to in \S~\ref{s-cb244_fit}; see \S~\ref{s-cb244}
for further details.  The signal-to-noise distribution for the source with mean $T \sim 20$ K is similar to that of the {\it Herschel} observations of
CB244. For this source, most of the pixels have
uncertainties dominated by the measurement noise, while the high $S /
N$ have uncertainties dominated by the calibration errors. However,
most pixels for the source with mean $T \sim 80$ K have uncertainties dominated
by the calibration errors. Therefore, the warmer source also provides
an interesting test regarding the importance of accounting for the
calibration uncertainties.

We applied both our Bayesian method and a $\chi^2$-based method to the
two simulated data sets. For the $\chi^2$ fits, we constrained the
best-fit temperature values to be between $1 < T < 100$ for the cooler
source and $1 < T < 300$ for the warmer source. We constrained the
column density to be positive and $\beta$ to lie within $-10 < \beta <
10$ for both sources. For each data point, we chose five random
independent initial guesses for $N,T,$ and $\beta$, and ran our
$\chi^2$ minimizer on each, keeping the value that minimized $\chi^2$
over the initial guesses. It is necessary to randomly initialize the
$\chi^2$ minimizing algorithm at multiple starting locations because
the algorithm did not always converge, or there may have been local
minima; this is mostly a problem for the low $S/N$ data points. For
the cooler source, we remove 104 data points for which the $\chi^2$
fits converged to a value on the boundary. For the warmer source, we
only exclude 2 data points which converged to the boundary.

\begin{figure*}
  \includegraphics[scale=0.5,angle=90]{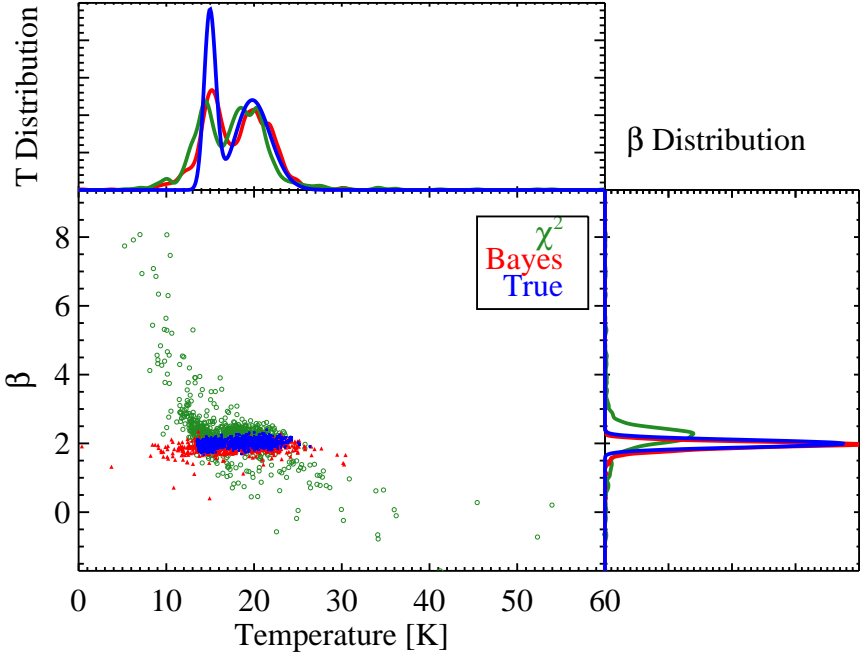}
  \caption{Distribution of the true values for $\beta$ and $T$ (blue
    squares) for a simulated source with mean $T \sim 20$ K, compared
    with a random draw from the posterior distribution using our
    Bayesian hierarchical model (red triangles) and $\chi^2$-based
    (green open circles) estimates. Also shown are the marginal
    distributions for temperature (top) and $\beta$ (right) for the
    true values (blue), Bayesian estimates (red), and $\chi^2$
    estimates (green). For clarity one data point with a
    $\chi^2$-based estimate of $T > 60$ K is excluded. The Bayesian
    estimates more accurately recover the true distribution of $\beta$
    and $T$ and their correlation, while the $\chi^2$-based estimates
    incorrectly show an anti-correlation and exhibit some bias in
    estimating the average $\beta$.}
  \label{f-temp20}
\end{figure*}

\begin{figure*}
  \includegraphics[scale=0.5,angle=90]{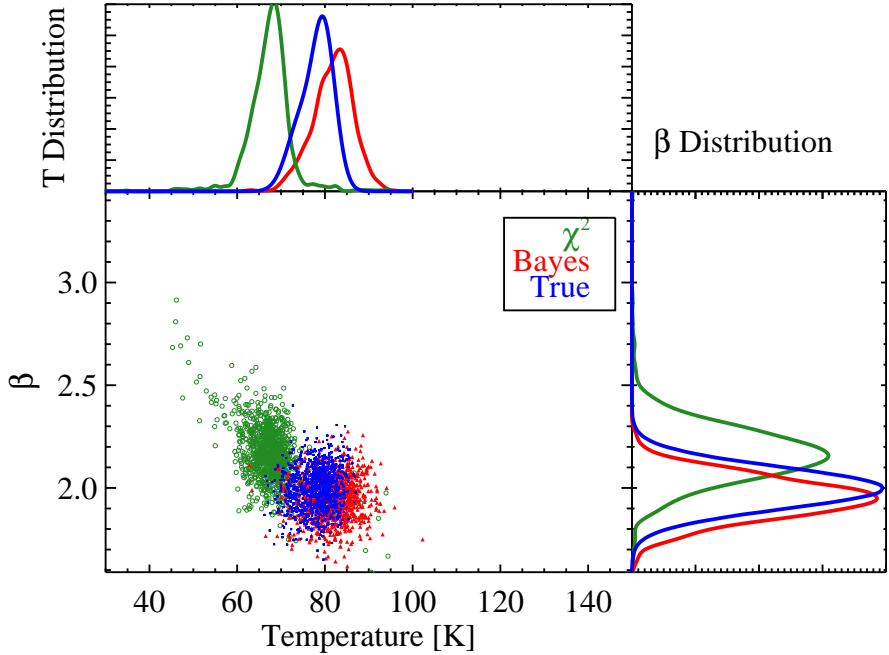}
  \caption{Same as Figure \ref{f-temp20}, but for a simulated source
    with mean $T \sim 80$ K. For clarity, two sources with
    $\chi^2$-based estimates of $T > 150$ K are excluded. As with the
    cooler source, our hierarchical Bayesian estimates provide a
    better reconstruction of the true distribution, although they
    exhibit a small bias in the estimated average temperature. In
    contrast, the $\chi^2$ estimates incorrectly imply a weak
    anti-correlation between $\beta$ and $T$, and exhibit some bias in
    the average values of $\beta$ and $T$.}
  \label{f-temp80}
\end{figure*}

In Figures \ref{f-temp20} and \ref{f-temp80} we show a random draw of
the values of $\beta$ and temperature for each pixel from their
posterior probability distribution for the source with mean $T \sim
20$ K and $T \sim 80$ K, respectively.  We plot a random draw of
$\beta$ and $T$ instead of the posterior median or mean because the
random draw of $\beta$ and $T$ more accurately reflects the spread of
the data in the $\beta$-$T$ plane. While the posterior median or mean
values provide better estimates of $\beta$ and $T$ for any individual
pixel, the distribtion of their values does not reflect as good of an
estimate of the distribution of $\beta$ and $T$, as they average over
the intrinsic variability in these quantities that is present in every
draw from the posterior distribution. In both Figures we also show the
distribution of the best-fit values obtained from minimizing $\chi^2$.
In agreement with \citet{Shettyetal09a}, we find that the
distributions for both $\beta$ and $T$ of the $\chi^2$-based estimates
are wider than the true distributions. Furthermore, {\it the
  distribution of the $\chi^2$-based estimates for $\beta$ and $T$
  portrays an anti-correlation between these two quantities, despite
  the fact that these two quantities are constructed to be positively
  correlated in our simulations}. This occurs because the
$\chi^2$-based estimate of the $\beta$--$T$ relationship is always
biased toward an anti-correlation, as the errors in $\beta$ and $T$
estimated by minimizing $\chi^2$ are anti-correlated. For this
simulation the errors on the $\chi^2$-based estimates are large, and
the magnitude of this bias is large enough to reverse the sign of the
correlation.

In Table \ref{t-compare} we compare the true values of the Spearman's
rank correlation between $\beta$ and $T$ for both simulations, denoted as $\rho$, with
those obtained from the $\chi^2$-based estimators and that obtained
from our hierarchical Bayesian method. In addition, in Table
\ref{t-compare} we compare the average values for $T$ and $\beta$, and
compare the values of the correlations and standard deviations of
the best-fit parameters; for the Bayesian estimates we use the values
of the correlations and standard deviations derived from the model
covariance matrix to investigate how well the model covariance matrix
recovers the true values. Unlike the $\chi^2$-based estimators, the best-fit estimates derived
using our hierarchical Bayesian method provide a more faithful reconstruction
of the intrinsic distribution of $\beta$ and $T$. In particular, the
Bayesian estimates recover the true positive correlation between $\beta$ and
$T$. 

\begin{deluxetable*}{ccccccc}[t]
  \tabletypesize{\scriptsize}
  \tablecaption{Comparison of Performance of $\chi^2$-based
    Estimators with Hierarchical Bayesian Estimates on Simulated Data
    with $T \sim 20$ K and $T \sim 80$ K\label{t-compare}}
  \tablewidth{0pt}
  \tablehead{
    & \multicolumn{3}{c}{Cooler Source} & 
    \multicolumn{3}{c}{Warmer Source} \\
    & \colhead{True}
    & \colhead{$\chi^2$}
    & \colhead{Bayes}
    & \colhead{True}
    & \colhead{$\chi^2$}
    & \colhead{Bayes}
  }
  \startdata
  $\rho$\tablenotemark{a} & 0.33 & -0.45 $\pm$ 0.03 & 0.23 $\pm$ 0.08 & 0.10 & -0.34
  $\pm$ 0.03 & 0.07 $\pm$ 0.02 \\
  $\bar{\beta}$\tablenotemark{b} & 2.0 & 2.26 $\pm$ 0.03 & 1.90 $\pm$ 0.07 & 2.0 & 2.16
  $\pm$ 0.01 & 1.96 $\pm$ 0.04 \\
  $\bar{T}$\tablenotemark{c} & 18.0 & 17.8 $\pm$ 0.17 & 18.3 $\pm$ 0.3 & 78.0 & 67.8
  $\pm$ 0.25 & 81.9 $\pm$ 0.18  \\
  Corr($\log N$,$\beta$)\tablenotemark{d} & -0.02 & -0.61 $\pm$ 0.02 & 0.00
  $\pm$ 0.06 & -0.01 & -0.07 $\pm$ 0.03 & 0.01 $\pm$ 0.04 \\
  Corr($\log T$,$\beta$)\tablenotemark{e} & 0.32 & -0.69 $\pm$ 0.02 & 0.38
  $\pm$ 0.03 & 0.09 & -0.62 $\pm$ 0.02 & 0.08 $\pm$ 0.03 \\
  Corr($\log N$,$\log T$)\tablenotemark{f} & -0.04 & 0.21 $\pm$ 0.03 &
  -0.06 $\pm$ 0.03 & -0.01 & 0.00 $\pm$ 0.03 & -0.01 $\pm$ 0.04 \\
  $\sigma$($\log N$)\tablenotemark{g} & 0.55 & 0.59 $\pm$ 0.01 & 0.58 $\pm$
  0.02 & 0.56 & 0.58 $\pm$ 0.01 & 0.60 $\pm$ 0.01 \\
  $\sigma$($\beta$)\tablenotemark{h} & 0.10 & 0.90 $\pm$ 0.02 & 0.111 $\pm$
  0.001 & 0.10 & 0.15 $\pm$ 0.01 & 0.11 $\pm$ 0.01 \\
  $\sigma$($\log T$)\tablenotemark{i} & 0.07 & 0.109 $\pm$ 0.002 & 0.083
  $\pm$ 0.001 & 0.02 & 0.04 $\pm$ 0.001 & 0.03 $\pm$ 0.001 \\
  \enddata
  
  \tablenotetext{a}{The value of Spearman's rank correlation
    coefficient between $\beta$ and $T$ for the simulated sources.}
  \tablenotetext{b}{The average value of $\beta$ for the simulated sources.}
  \tablenotetext{c}{The average value of temperature for the simulated
    sources.}
  \tablenotetext{d}{The model value for the correlation between $\log N$
    and $\beta$}.
  \tablenotetext{e}{The model value for the correlation between $\log T$
    and $\beta$}.
  \tablenotetext{f}{The model value for the correlation between $\log N$
    and $\log T$}.
  \tablenotetext{g}{The model value for the standard deviation in
    $\log N$.}
  \tablenotetext{h}{The model value for the standard deviation in
    $\beta$.}
  \tablenotetext{i}{The model value for the standard deviation in
    $\log T$.}

\end{deluxetable*}

Our Bayesian method correctly recovers the true values of the means
and correlation coefficients within the uncertainties. The only
exception to this is the mean temperature for the warmer source, for
which our Bayesian method ovestimates the true value by $\approx 4$ K,
a difference of about $5\%$. This bias is most likely caused by our
incorrect assumption of a Student's $t$-distribution for the intrinsic
distribution of $(\log N,\log T,\beta)$. However, this bias is very
small, and in general our Bayesian approach recovers the correct
values, illustrating the robustness of our method to inaccuracies in
the assumed distribution. In contrast, the distribution derived from
the $\chi^2$-based estimates is biased and leads to incorrect
conclusions. In addition to a spurious anti-correlation between
$\beta$ and $T$, the $\chi^2$-based estimates also exhibit biases in
the estimated mean values of $\beta$ and $T$. The $\chi^2$-based
method correctly recovers the mean temperature for the cooler source,
but overestimates the mean value of $\beta$ by $\approx 13\%$. The
bias in the mean values is more noticeable for the warmer source,
where the fluxes only sample the Rayleigh-Jeans regime.  Here, the
mean $\beta$ is overestimated by $\approx 8\%$ while the mean
temperature is underestimated by $\approx 15\%$. The most likely
source of this bias is the calibration errors, which the $\chi^2$
method does not account for. In contrast, our Bayesian approach
corrects for the calibration errors, and incorporates their
contribution to the uncertainty in the parameters estimates.

Our hierarchical Bayesian method also does a better job of recovering
the true values of $N,\beta,$ and $T$ for individual pixels. For the
cooler source the median absolute values of the error in the Bayesian
estimates for $\log N$, $\beta$, and $T$ are 0.05, 0.11, and 0.59,
respectively. For the $\chi^2$ estimates, these values are 0.08, 0.25,
and 1.08 for $\log N, \beta,$ and $T$. The Bayesian posterior median
estimates do a factor of $\approx 2$ better that the $\chi^2$
estimates for the source with mean $T \sim 20$ K. For the source with
mean $T \sim 80$ the Bayesian fits for individual pixels also did
better, but this time the error in the $\chi^2$ estimates are
dominated by the bias caused by the unaccounted-for calibration
errors.

\begin{figure}
  \includegraphics[scale=0.3,angle=90]{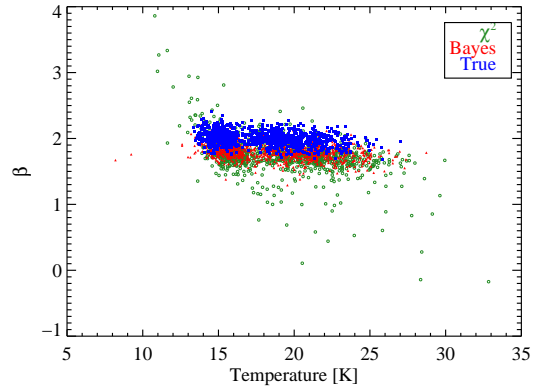}
  \caption{Distribution of true values of $\beta$ and $T$ (blue square
    points) for a simulated data set with an intrinsic
    anti-correlation enforced, compared with the distribution of the
    $\chi^2$-based estimates (green open circles) and a random draw
    from the posterior distribution under our hierarchical Bayesian
    model (red triangles). Both methods are able to recover the
    anti-correlation, but the $\chi^2$-based estimates produce a
    stronger and more extended \Tbeta\ anti-correlation compared to
    the true distribution.}
  \label{f-anticorr}
\end{figure}

As a sanity check, we also show that our hierarchical Bayesian method
recovers an anti-correlation when one exists. We simulate values of
$\beta, T,$ and $N$ in exactly the same manner as for the source with
mean $T \sim 20$ above, but this time enforce an anti-correlation by
using a value of $B = -0.5$. We then apply our hierarchical Bayesian
method to the simulated data, and obtain estimates via
$\chi^2$-minimization. Figure \ref{f-anticorr} compares the
results. For this simulation, the true value of the Spearman's rank
correlation coefficient between $\beta$ and $T$ is $\rho = -0.28$. The
value inferred from our Hierarchical Bayesian estimate is $\rho =
-0.27 \pm 0.02$, while the value inferred from the $\chi^2$-based
estimates is $\rho = -0.38 \pm 0.03$. Both the Bayesian and $\chi^2$
method recover the anti-correlation.  However, as expected the
$\chi^2$ estimates produces an artificially stronger and more extended
\Tbeta\ anti-correlation (see Section 2.1).

As a final test of our method we generate fluxes from an idealized
simple model of a starless core.  We construct a projected core
similar to Model 1 from \citet{Shettyetal09b}, based on work by
\citet{evans01}.  In this model, $N$ increases linearly toward the
central regions, ranging from $2\times 10^{21}$ to $1.25\times
10^{22}$ \cmt.  Correspondingly, $T$ decreases linearly, from 12 to 8
K.  Lastly, $\beta$ decreases from 2.6 to 1.8. Our choice of the
quantitative values for $T$, $N$, and beta are motivated by the
results we obtain from CB244 (see \S~\ref{s-cb244_fit}), and are
generally consistent with the trends discussed by
\citep{evans01}. This model provides a test of how our method performs
in the limiting scenario where the values of $N,T,$ and $\beta$ all
lie on simple deterministic curves which do not have any intrinsic
dispersion.  We do not consider this model to be realistic, as a real
cloud is unlikely to be spherically symmetric with the line-of-sight
averaged values of $N$, $T$, and $\beta$ lying along deterministic
curves.  However, it does provide a useful test of our method as this
model may violate the assumption that the covariance matrix of $(\log
N, \log T,\beta)$ is positive definite.

For this model we simulate flux values as before, and execute both our
hierarchical Bayesian method and the $\chi^2$ fit. The true
correlation coefficients are ${\rm Corr}(\log N,\beta) = -0.98$, ${\rm
  Corr}(\log N, \log T) = -0.97$, and ${\rm Corr}(\log T, \beta) =
0.99$. Our hierarchical Bayesian method returns values of ${\rm
  Corr}(\log N,\beta) = -0.95 \pm 0.03$, ${\rm Corr}(\log N, \log T) =
-0.98 \pm 0.001$, and ${\rm Corr}(\log T, \beta) = 0.88 \pm 0.04$,
while the $\chi^2$-based estimates infer ${\rm Corr}(\log N,\beta) =
-0.41$, ${\rm Corr}(\log N, \log T) = -0.31$, and ${\rm Corr}(\log T,
\beta) = -0.71$.  Our hierarchical Bayesian method again outperforms
the $\chi^2$ approach for this test, although it does not perform as
well as in previous tests. In addition, as with the other tests the
$\chi^2$ method incorrectly produces an anti-correlation between
$\beta$ and $T$.

For this test, convergence of our MCMC sampler proves to be extremely
slow due to the strong dependencies among the values of $N, \beta,$
and $T$ and $\Sigma$. Because the correlations for this test are $|r|
\approx 1$, $N,\beta,$ and $T$ for each pixel are very precisely
determined from $\mu$ and $\Sigma$. However, given $N, \beta,$ and $T$
for each pixel, $\mu$ and $\Sigma$ are very precisely determined. This
dependency results in very slow convergence of our MCMC sampler, and
the decreased performance is likely due to this lack of
convergence. While this does not affect the qualitative results,
convergence should be carefully monitored when there is evidence that
the correlations from $\Sigma$ converge to 1 or -1. Future additions
to our MCMC sampler will improve convergence for cases where the
correlations among $N, \beta,$ and $T$ are extremely tight.

\section{Application: Dust in star forming Bok globule CB244}\label{s-obssec}

In this section we discuss the application of our Hierarchical
Bayesian method to the Bok globule CB244. The purpose of this
application is to illustrate how our method performs on real data,
to compare the results from our method with those obtained from
$\chi^2$ minimization, and to illustrate what type of conclusions might be derived
from our method as compared to traditional non-hierarchical
methods. The application of our method to CB244 is valuable as an illustration because real
data is often subject to further complications beyond the simple 
idealized simulations that we have performed, and can include a number
of systematics that are not captured in our statistical model. A full
treatment of all data systematics is beyond the scope of this
methodology-focused article, but will be addressed in future work
when we apply our method to similar astronomical sources. 

\subsection{Observations}\label{s-cb244}

CB244 is an isolated, low--mass, star--forming molecular cloud located
at a distance of $\sim 200$~pc.  It contains two {\it Herschel}
emission peaks, one associated with a Class 0 protostar and one
associated with a starless core \citep{Stutzetal10}.  Because of the
relative simple geometry of such sources, Bok globules are excellent
targets to study the processes taking place in the dense ISM, free
from complications arrising in more clustered environments.  The {\it
Herschel} CB244 data were acquired as part of the Guaranteed Time Key
Programme ``Earliest Phases of Star--formation'' \citep[EPoS; P.I.\
O. Krause, e.g.,;][]{beuther10,henning10,linz10,Stutzetal10} as part
of the Science Demonstration Program.  The sources in this program
were selected to be in relatively isolated regions in order to
minimize the effects from uncertain background subtraction. These data
were first presented in \citep{Stutzetal10}.  The submm SCUBA
870~\micron\ and the IRAM 1.3~mm ground--based data were presented in
\citet{laun10}.  For our purposes of temperature mapping, the data
reduction has been updated; here we present a brief outline of the
{\it Herschel} processing.  See Launhardt et al., (2012, in prep) for
further details.

The {\it Herschel} \citep{pil10} data for CB244 were observed with the
Photodetector Array Camera and Spectrometer \citep[PACS;][]{pog10} and
Spectral and Photometric Imaging Receiver \citep[SPIRE;][]{griff10}
The PACS 100 \micron\ and 160 \micron\ observations were carried out
on December 30, 2009.  Two orthogonal scan maps were acquired using a
scan speed of 20$\arcsec$/s with scan leg lengths of 9$\arcmin$.  The
AOR ID's for these observations are 134218869(4,5).  The two
wavelengths were processed in an identical fashion.  The level 1 data
were processed using HIPE v.\ 6.0.1196.  Final level 2 maps were
produced using Scanamorphos (Roussel et al., 2011, in prep.), using
the ``galactic'' option.  These data were processed including the
non--zero--acceleration telescope turn--around data.  The SPIRE 250,
350, and 500 \micron\ observations were obtained on October 20, 2009.
Two 9$\arcmin$ scan legs were obtained at the nominal scan speed of
30$\arcsec$/s.  The AOR ID for these data is 1342199366.  These data
were processed up to level 1 with HIPE 5.0.1892. The level 2 maps were
processed using Scanamorphos v.9 (Roussel et al., 2011, in prep.).

The calibrated dust emission maps were then processed in an identical
fashion to the data presented in Launhardt et al., (2012, in prep).
We briefly summarize the steps used here.  First, the data are
re-zeroed using a method similar to that applied to {\it Spitzer MIPS}
images in \citet{Stutzetal09}. In order to caculate the DC-level
offset from the data we identified a $4' \times 4'$ emission--free
region that appears `dark' in the {\it Herschel} maps. In addition, we
require that this region is in or near a region which is relatively
free from $^{12}{\rm CO} (2$--$1)$ emission (Launhardt et al. 2012, in
prep). The same spatial region was used for all five {\it Herschel}
maps. For each band, we then calculate the representative flux level
in the region by implementing an iterative Gaussian function fitting
and sigma-clipping scheme to the pixel value distribution at each
wavelength. The mean value of the best--fit Gaussian function is
subtracted from each image, while the standard deviation is used to
estimate the noise levels.  If the noise in the image is Gaussian,
then the distribution of measured flux values for pixels with true
flux near the DC level will also be Gaussian. In this case, the mean
of the best-fit Gaussian function provides an estimate of the DC
level, while the standard deviation of the best-fit Gaussian function
provides an estimate of the noise amplitude. The estimated DC levels
and noise levels for each {\it Herschel} map are reported in Table
\ref{t-dclevels}.

\begin{deluxetable*}{cccccc}[t]
  \tabletypesize{\scriptsize}
  \tablecaption{Estimated DC-levels, Noise Amplitudes, and Calibration Errors for
    \emph{Herschel} maps of CB244 \label{t-dclevels}}
  \tablewidth{0pt}
  \tablehead{
    & \colhead{100\micron}
    & \colhead{160\micron}
    & \colhead{250\micron}
    & \colhead{350\micron}
    & \colhead{500\micron}
  }
  \startdata
  DC-level\tablenotemark{a} & 1298 & 510.0 & 298.9 & 169.6 & 77.19 \\
  Noise Amplitude\tablenotemark{b} & 39.52 & 75.00 & 26.54 & 13.44 & 7.206 \\
  Estimated Calibration Errors\tablenotemark{c} & $0.93 \pm 0.14$ &
  $0.90 \pm 0.09$ & $1.07 \pm 0.10$ & $ 1.18 \pm 0.14$ & $1.26 \pm 0.19$ \\
  \enddata
  
  \tablenotetext{a}{DC-level in units of $\mu {\rm Jy} / \arcsec^2$,
    estimated according to the procedure described in
    \S~\ref{s-cb244}}

  \tablenotetext{b}{Standard deviation in the additive noise in units of $\mu {\rm Jy} / \arcsec^2$,
    estimated according to the procedure described in
    \S~\ref{s-cb244}}

  \tablenotetext{c}{The posterior median and standard deviation for
    the calibration errors. The calibration
    errors are \emph{a priori} assumed to be log-normally distributed
    with a geometric mean of unity and an uncertainty of 15\%.}

\end{deluxetable*}

A pointing correction is also applied to the PACS images relative the
the MIPS 24~\micron\ data of the same field; in the case of CB244 this
correction is small (of order 1$\arcsec$) compared to the SPIRE
500~\micron\ beam.  We apply the recent {\it Herschel} calibration
correction factors to the data.  The 100 to 500~\micron\ data are then
converted to common units of Jy/$\arcsec^2$.  The data (including the
sub--mm ground--based observations) are then convolved to the limiting
beam, in this case the SPIRE 500~\micron\ beam.  We use the
\citet{aniano11} circularized convolution kernels for the 100 to
350~\micron\ data, and a Gaussian beam approximation for the
ground--based sub--mm data.  Finally, the data are re-gridded to a
common coordinate system and a pixel scale of 10$\arcsec$.

\subsection{Fitting results}\label{s-cb244_fit}

\begin{figure}
  \includegraphics[scale=0.5,angle=-90]{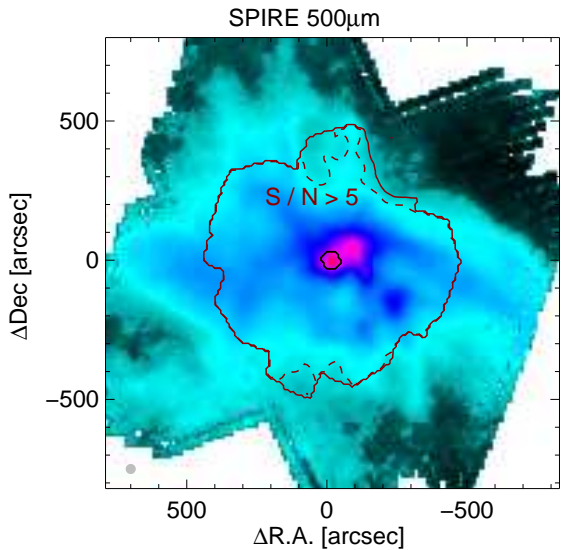}
 \caption{Coverage of the CB244 pixels for which we analyzed using our
   hierarchical Bayesian method, compared with the SPIRE 500~\micron\
   map. We analyzed pixels that had coverage in all five
   \emph{Herschel} bands and a mean $S/N$ over the \emph{Herschel}
   bands of $\langle S/N \rangle > 2$. In addition, we omitted the
   protostar in the center of the image from our analysis. The grey
   circle in the lower left corner illustrates the size of the SPIRE
   500~\micron\ beam.} 
 \label{f-coverage}
\end{figure}

In applying our Bayesian method to the CB244 dataset, we limit our
analysis to those pixels containing fluxes in at least each of the five {\it
  Herschel} bands, since the SPIRE images have more coverage than the
PACS images.  In addition, to minimize the impact of uncertainties in the
estimated DC-level offset, we limit our analysis to those pixels
having $S/N > 2$ as averaged over the five {\it Herschel} bands, for
which the formal statistical error in the DC-level offset is negligible. The coverage of
pixels that we analyze using our hierarchical Bayesian method is shown
on the 500~\micron\ map in Figure \ref{f-coverage}.

We conservatively assume calibration uncertainties of $15\%$ for each
of the {\it Herschel} bands, and calibration uncertainties of $30\%$
and $20\%$ for the 870 \micron\ and $1.3$~mm bands, respectively. The
values we adopt are larger than the official \emph{Herschel}  Science
Center (HSC) values, and our reasons are as follows. 

For the PACS instrument, the point-source calibration uncertainty is
3\% and 5\% at 100 \micron\ and 160 \micron, respectively. However, we
have used versions of the Launhardt et al. (2012, in prep) data
reduced with Scanamorphos, a pipeline that is better suited for the
analysis of extended emssion compared to high-pass
filtering. High-pass filtering is another
common Herschel image analysis technique that removes unknown levels of
extended emission and is thus better suited for point-source analysis.
We note that the high-pass filtering technique removes 
$1/f$ noise more efficiently than other map processing techniques;
therefore our Scanamophos maps may have elevated noise levels by
comparison.  

Because we are most interested in investigating the extended emission, we conclude that
reduction techniques that remove extended emission levels, namely
high-pass filtering, are not robust for out scientific goals even when
they deliver data products with reduced $1/f$ noise. The main remaining
uncertainties are most likely caused by beam convolution effects
(e.g., imperfect kernels) and possibly color corrections.  These
uncertainties are very hard to quantify.  Therefore our strategy is to
adopt an inflated calibration uncertainty, meant to represent multiple independent source of uncertainty:
extended emission calibration uncertainties, the uncertainties
introduced by beam convolution, $1/f$ noise, and other unidentified
effects. Thus the final calibration uncertainties we use are 15\% and
are conservative compared to the HSC recomended point-source
calibration uncertainties.

For SPIRE the final HSC recommended calibration uncertainty is
$7\%$. However, the calibration for extended sources is performed by
multiplying the calibration for point sources by a correction
factor (see Section 5.2.8 of the SPIRE Observers' Manual\footnote{\url{http://herschel.esac.esa.int/Docs/SPIRE/html/spire\_om.html}}). No
additional extended source calibration uncertainty is 
discussed to our knowledge, and thus we also assume 15\% uncertainty
on the calibration for the SPIRE maps. These values are conservative compared to
the HSC recomended values for calibration uncertainties.

We employed the color corrections reported in the \emph{Herschel/SPIRE}
and \emph{Herschel/PACS} observer's manuals. In general, the color
corrections are small. Following \citet{Schneeetal10}, we ignore the
color correction for the $\lambda = $ 870 \micron\ data, and impose a
color correction to the 1.3~mm data by modifying the effective
wavelength to be $\lambda = 1.1$~mm.

Observed fluxes from the centrally-heated protostellar region are
affected by line-of-sight temperature variations, and possibly even
optically thick dust.  Due to these systematic uncertainties, the
estimated protostellar values of \N, \bt, and \T\ through
Eq. \ref{eq-modbbody} are likely to be highly erroneous, and may even
introduce biases into other inferred quantities, such as the mean and
covariance of $\log N(H), \log T$, and $\beta$ of the whole sample, as
well as the calibration uncertainties.  Accordingly, in our analysis
we omit the pixels corresponding to the prototstar.
 
We also estimate $N(H),T,$ and $\beta$ based on minimizing
$\chi^2$. Because the $\chi^2$-based estimates are unstable at low
$S/N$, we limit our analysis to those pixels for which the average
$S/N$ over the {\it Herschel} bands is $\langle S/N \rangle > 5$.

\begin{figure*}
  \includegraphics[scale=0.35,angle=90]{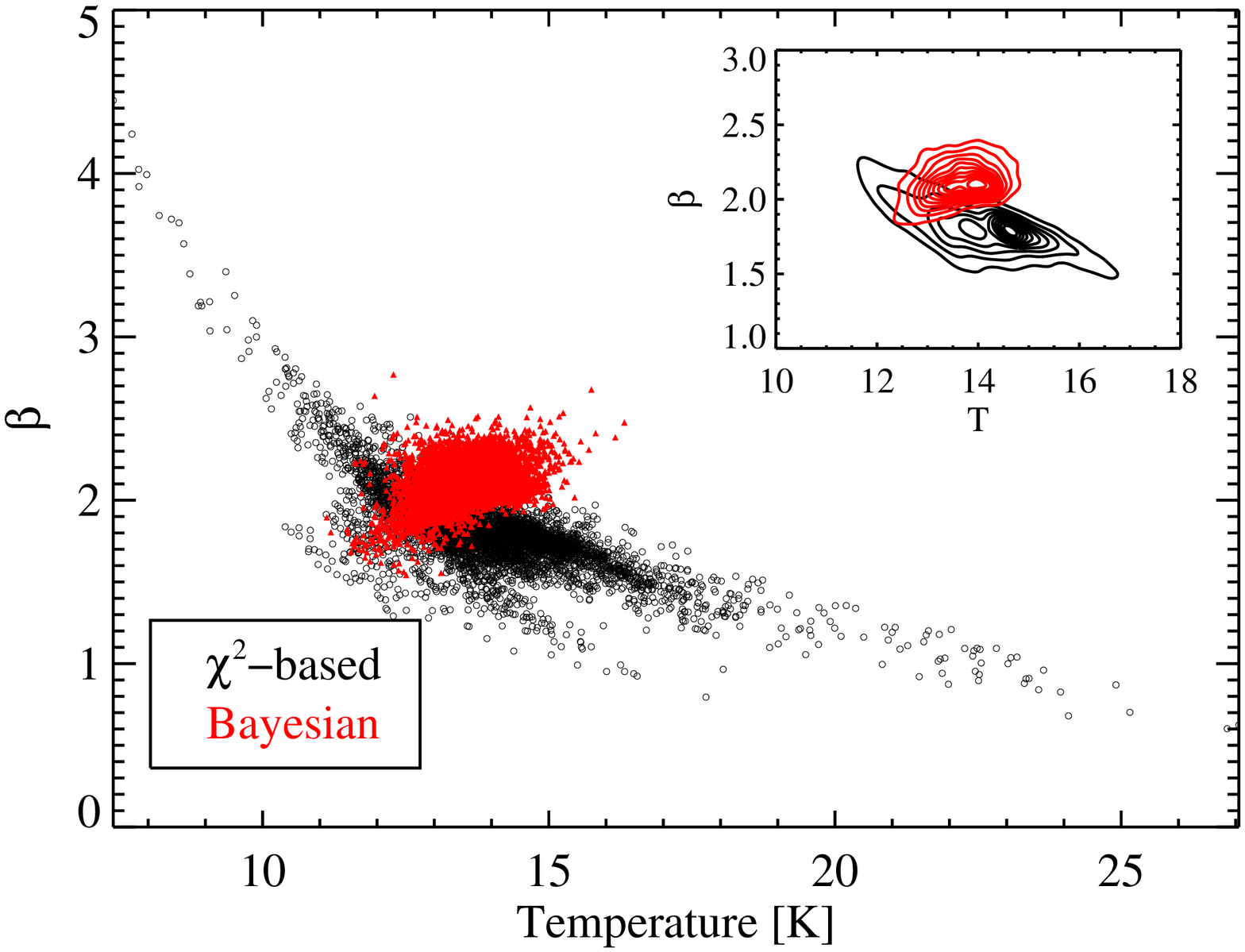}
  \includegraphics[scale=0.35,angle=90]{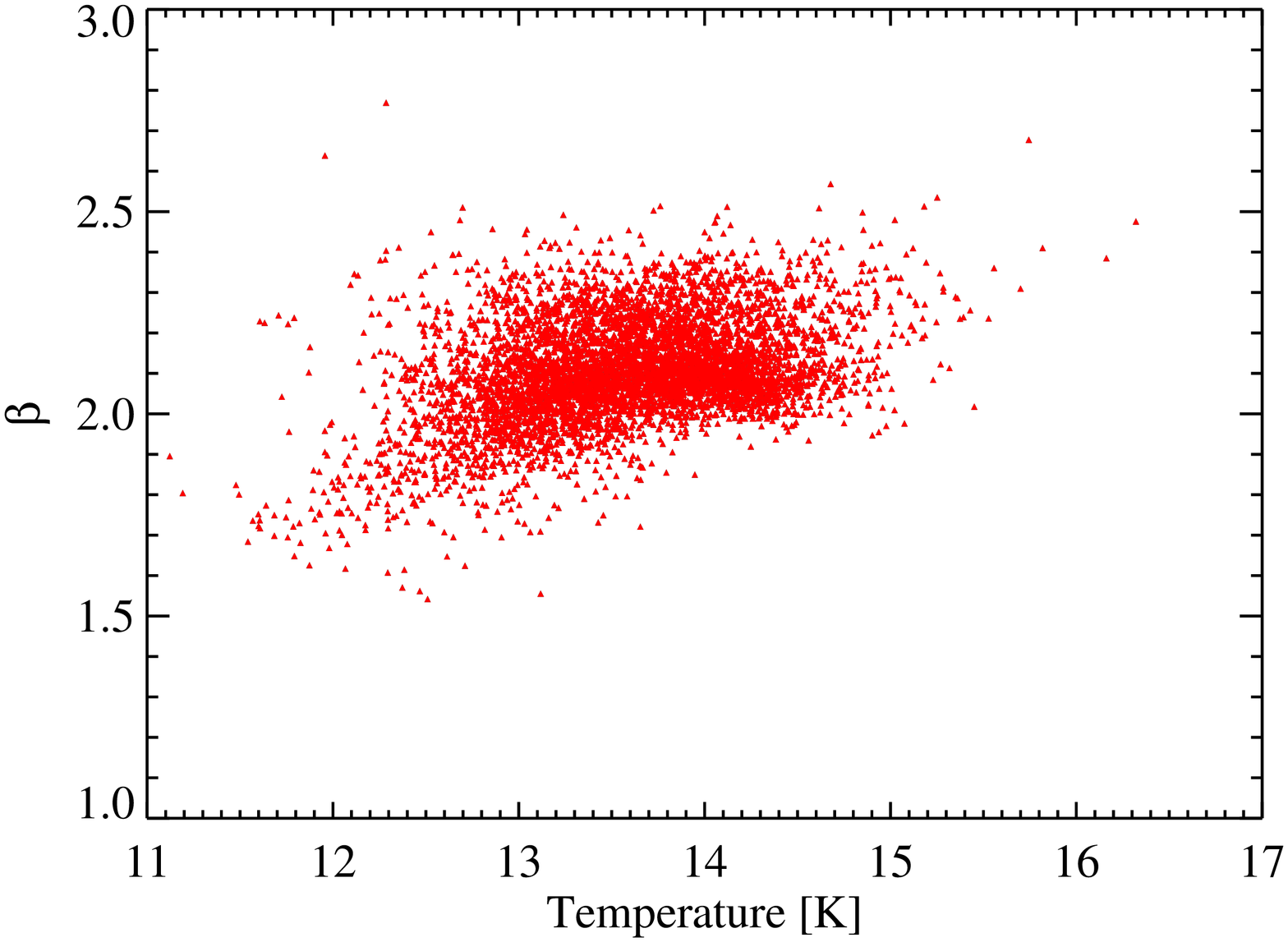}
  \caption{The left panel shows the distribution of $\beta$ and $T$
    for CB244 from minimizing $\chi^2$ (black open circles) and the
    random draw from the posterior distribution under our hierarchical
    Bayesian model (red triangles). The inset provides a close-up of
    the density of the distribution, while the right panel shows a
    close-up of the hierarchical Bayesian values. As expected, the
    $\chi^2$-based estimates display an anti-correlation. However, the
    Bayesian estimates show a weak positive correlation, and there is
    a large range in $\beta$ at fixed $T$.}
 \label{f-cb244_betat}
\end{figure*}

The derived relationship between $\beta$ and temperature for both the
Bayesian and $\chi^2$-based estimates are shown in Figure
\ref{f-cb244_betat}.  The $\chi^2$-based estimates suggest a strong
anti-correlation between $\beta$ and $T$, as expected.  However, the
Bayesian analysis finds that $\beta$ weakly increases as $T$
increases, which is the opposite trend compared to the $\chi^2$-based
estimates. The Spearman's rank correlation for the $\chi^2$-based
estimates is $\rho = -0.68 \pm 0.01$, while for the Bayesian estimates
it is $\rho = 0.33 \pm 0.04$. The correlation between $\beta$ and $T$
is rather weak, and while $\beta$ tends to increase with $T$ in the
mean, there is a large scatter in $\beta$ at a given temperature.

\begin{figure*}
  \includegraphics[scale=0.3,angle=0]{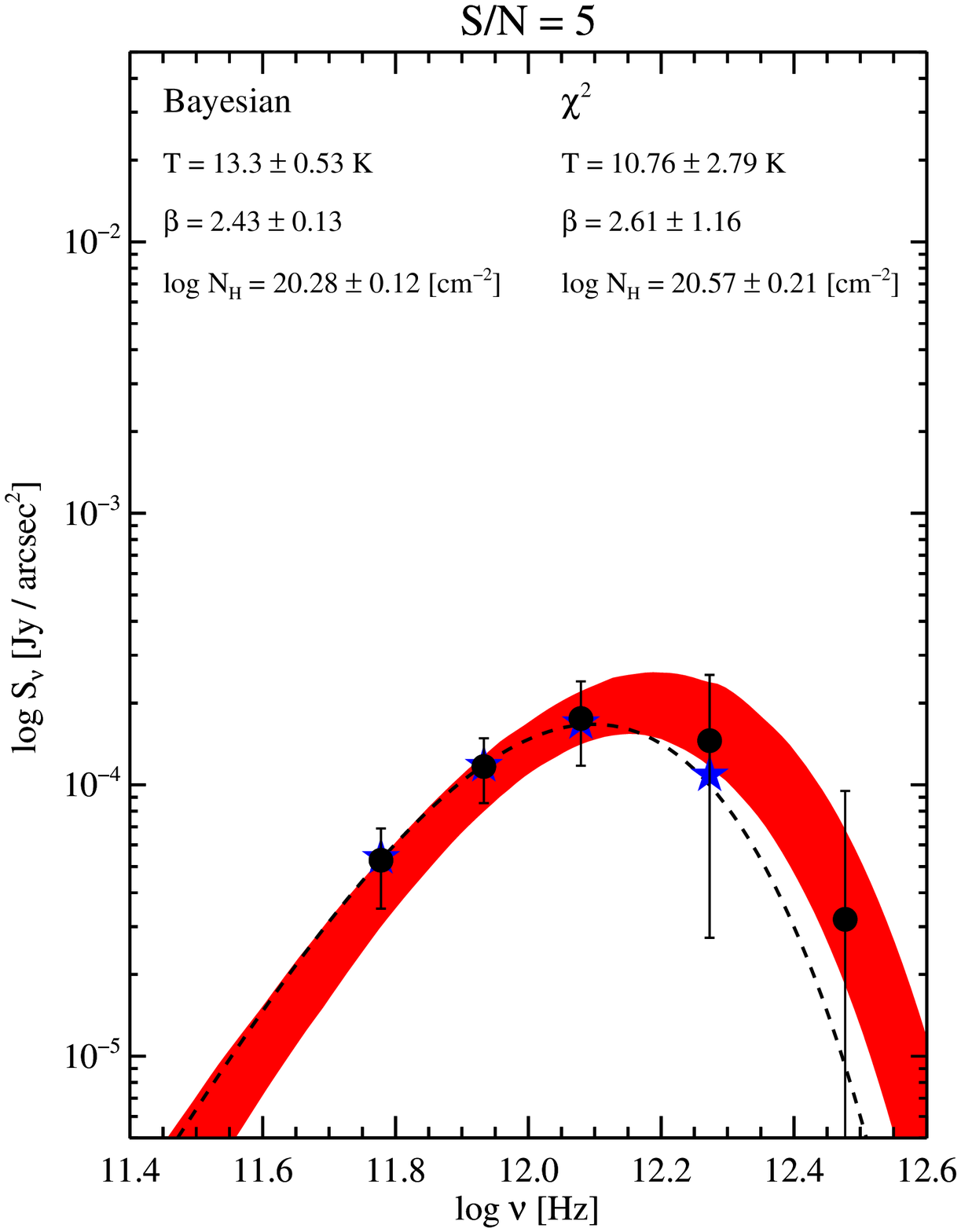}
  \includegraphics[scale=0.3,angle=0]{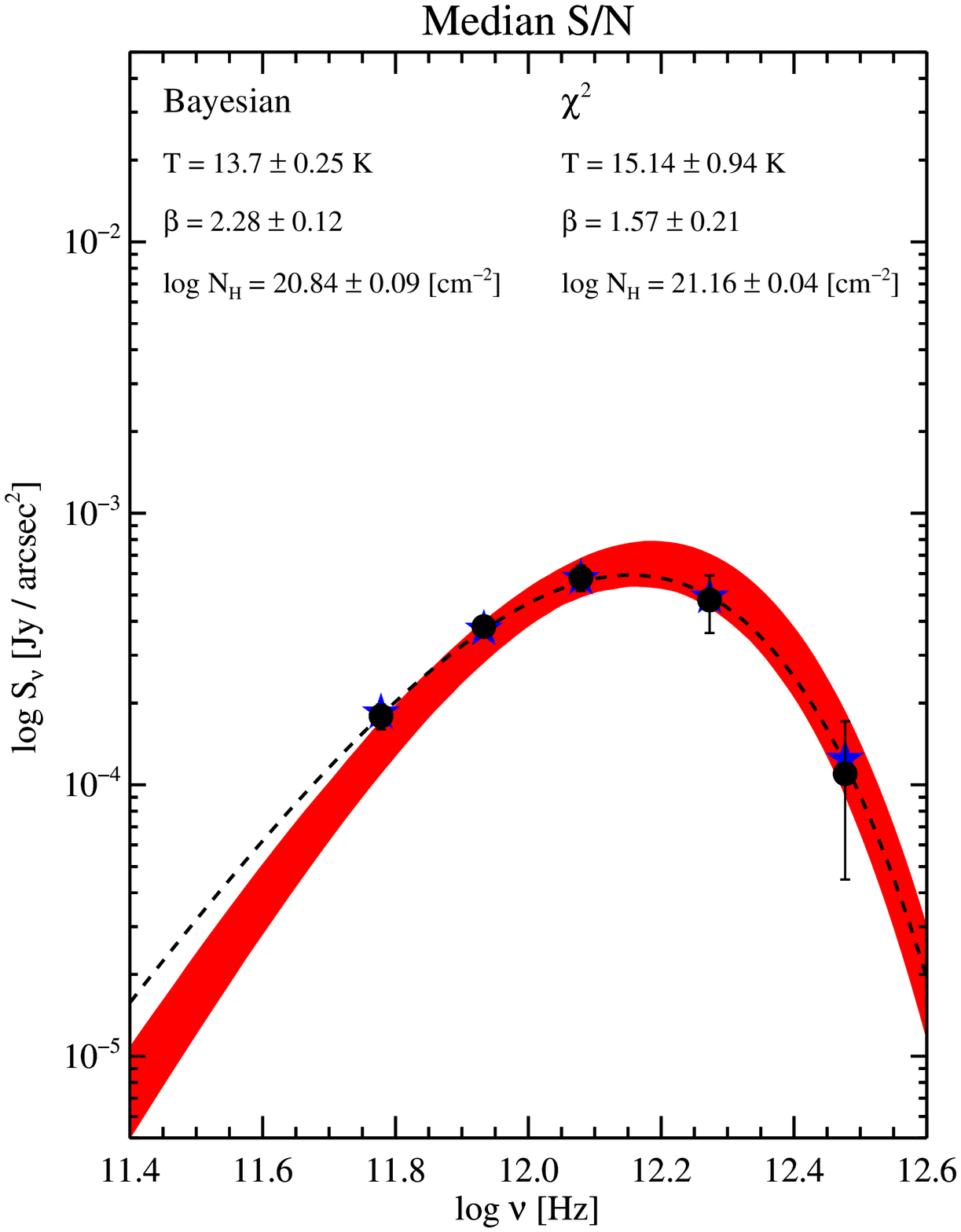}
  \includegraphics[scale=0.3,angle=0]{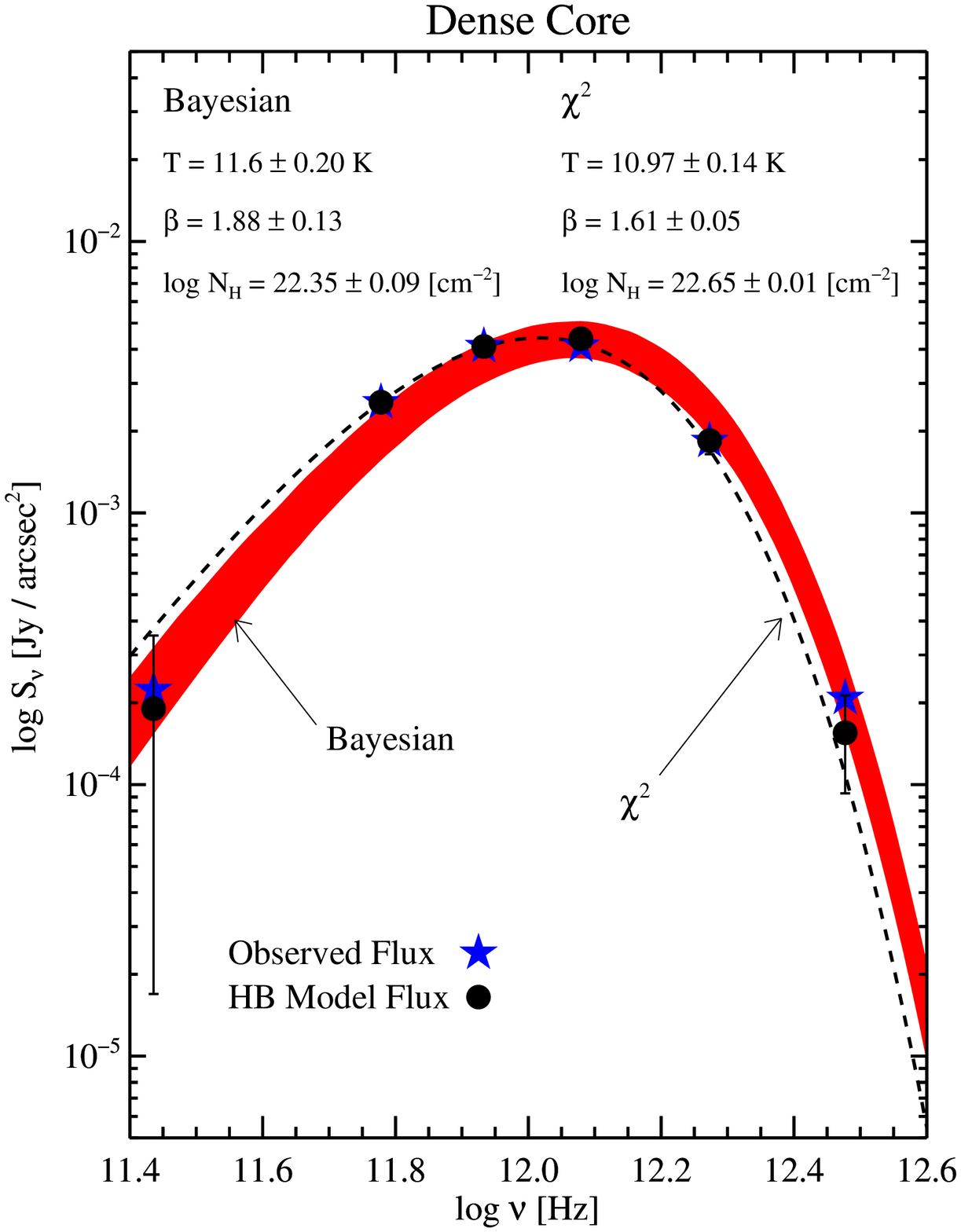}
  \caption{Measured fluxes (blue stars) for the pixel with the highest
   estimated column density in the prestellar core (right), a pixel
   with average Herschel $S/N$ similar to the median value (center),
   and the pixel with average Herschel $S/N = 5$ (left), which defines
   the lower limit of our $S/N$ cut for the $\chi^2$ estimates. The
   $100\mu$m flux measurement is missing from the left panel because
   its value is negative. The best-fit greybody SEDs derived from the
   $\chi^2$ estimates are shown with a dashed black line, while the
   red regions contain $95\%$ of the posterior probability for the
   greybody SEDs derived from our hierarchical Bayesian method. The
   measured fluxes are compared with the values that are predicted
   from our Bayesian model (black circles), with the error bars
   containing $95\%$ of the posterior probability on the measued
   SED. The fluxes and their error bars predicted from our Bayesian
   model differ from the model greybody SEDs in that they also include
   the effects of the calibration error and noise, and thus it is the
   green circles that should be compared with the measured data and
   not the red region. The actual measured values of the flux fall
   within the range expected from our Bayesian model, and therefore
   our model is consistent with the measured data.}
 \label{f-cb244_seds}
\end{figure*}

In Figure \ref{f-cb244_seds} we show the SED for the pixel in the
prestellar core with highest $N(H)$, a pixel with $\langle S/N
\rangle$ similar to the median value, and a pixel with $\langle S/N
\rangle = 5$.  For the prestellar core we find $T = 11.6 \pm 0.2$ K,
$\beta = 1.88 \pm 0.13$, and $\log N(H) = 22.65 \pm 0.01\ {\rm
cm^{-2}}$. For the prestellar core, the $\chi^2$ estimates are $T =
10.97 \pm 0.14$ K, $\beta = 1.61 \pm 0.05$, and $\log N(H) = 22.35 \pm
0.09\ {\rm cm^{-2}}$. The SEDs are compared with the range of greybody
models that contain $95\%$ of the posterior probability. In addition,
we show the SED derived from the $\chi^2$ estimates. Note that the
Bayesian greybody SED models defined by the red region do not include
the contribution from the calibration uncertainties, and thus it is
not appropriate to compare them directly to the data.

In order to assess the quality of fit, we compare the measured fluxes
with those predicted by our Bayesian method for each of the three SEDs
shown in Figure \ref{f-cb244_seds}. This is called a ``posterior
predictive check'' \citep{rubin81,rubin84,gelman96}, and is commonly
used to assess the goodness-of-fit of a Bayesian model. For each
random draw from our MCMC sampler, we simulate a flux value at each
value of $\nu$, incorporating the calibration error and measurement
noise.  Through this process we build up a distribution of predicted
fluxes incorporating our uncertainty in the derived parameters.  The
advantage of this approach is that we compare the actual measured
fluxes (which are considered fixed and known) to a distribution of
predicted fluxes, instead of simply comparing the measured fluxes at
each wavelength to a single best-fit SED. Because the predicted fluxes
also have the calibration errors and noise folded in, they are the
appropriate quantity to compare to the measured fluxes to test the
quality of the fit, and not the model greybody SEDs. Figure
\ref{f-cb244_seds} also compares the measured fluxes with the ranges
containing $95\%$ of the values predicted from our Bayesian approach.
In all cases the measured values fall within this range, showing that
the Bayesian results are consistent with the measured data, and
therefore provide an acceptable fit.

We also checked the derived calibration errors and their uncertainties,
in order to ensure that the derived calibrations are consistent with
those obtained from the data reduction; i.e., that the calibration
errors are consistent with $\delta_j = 1$. The estimated calibration
errors and their uncertainties are also listed in Table
\ref{t-dclevels}. The estimated calibration errors are consistent with
unity, implying that we do not find any evidence for significant
deviations from the calibrations performed in the data reduction which
might be indicative of data systematics or model
mis-specification. Moreover, the posterior uncertainties in the derived
calibration errors are similar to the \emph{a priori} assumed values
of 15\%, indicating that essentially all of the information from the
calibration errors comes from the prior that we have placed on
them. Because of this, the fact that our method incorporates the
calibration errors should not be interpreted as a recalibration of
the data. Rather, we have included the calibration errors as nuisance
parameters which are identified as an additional source of measurement
error. Indeed, it is not the absolute calibration which is included in
our statistical model, but rather the error in the
calibration. The posterior probability distributions that we obtain
from our MCMC method average over the unknown calibration errors, thus
ensuring that uncertainties in the calibration are also reflected in
the uncertainties in the derived greybody parameters, as well as
reflected in the estimated means and correlations (also see the discussion
in \S~\ref{s-cb244_systematics}).

As an additional test, we perform cross-validation to compare the
hierarchical Bayesian estimates with the $\chi^2$-based ones. For this
test, we randomly remove $10\%$ of our photometry, and then refit the
remaining $90\%$ of the CB244 data. The resulting estimates of
$N(H),\beta,$ and $T$ were then used to predict the flux values for
the $10\%$ of the data that were left out. If the hierarchical
Bayesian estimates provide a better description of the SED, then they
should do a better job of predicting the data that is omitted from the
fit. As a measure of the error in the flux, we use the absolute value
of the difference between the measured flux and the model flux,
divided by the standard deviation of the noise in that band,
$\sigma_j$. For our Bayesian estimates, the median of this error is
$0.55$, while the median for the $\chi^2$ estimates is $0.90$. The
hierarchical Bayesian estimates did a factor of $\sim 2$ better than
the $\chi^2$ estimates in predicting the flux values for data that is
omitted from the fit. This result suggests that the Bayesian estimates
provide a better description of the SED, and are therefore more
reliable than the $\chi^2$ ones.

\begin{figure*}
  \includegraphics[scale=0.5, angle=-90, bb= 88 41 524 500]{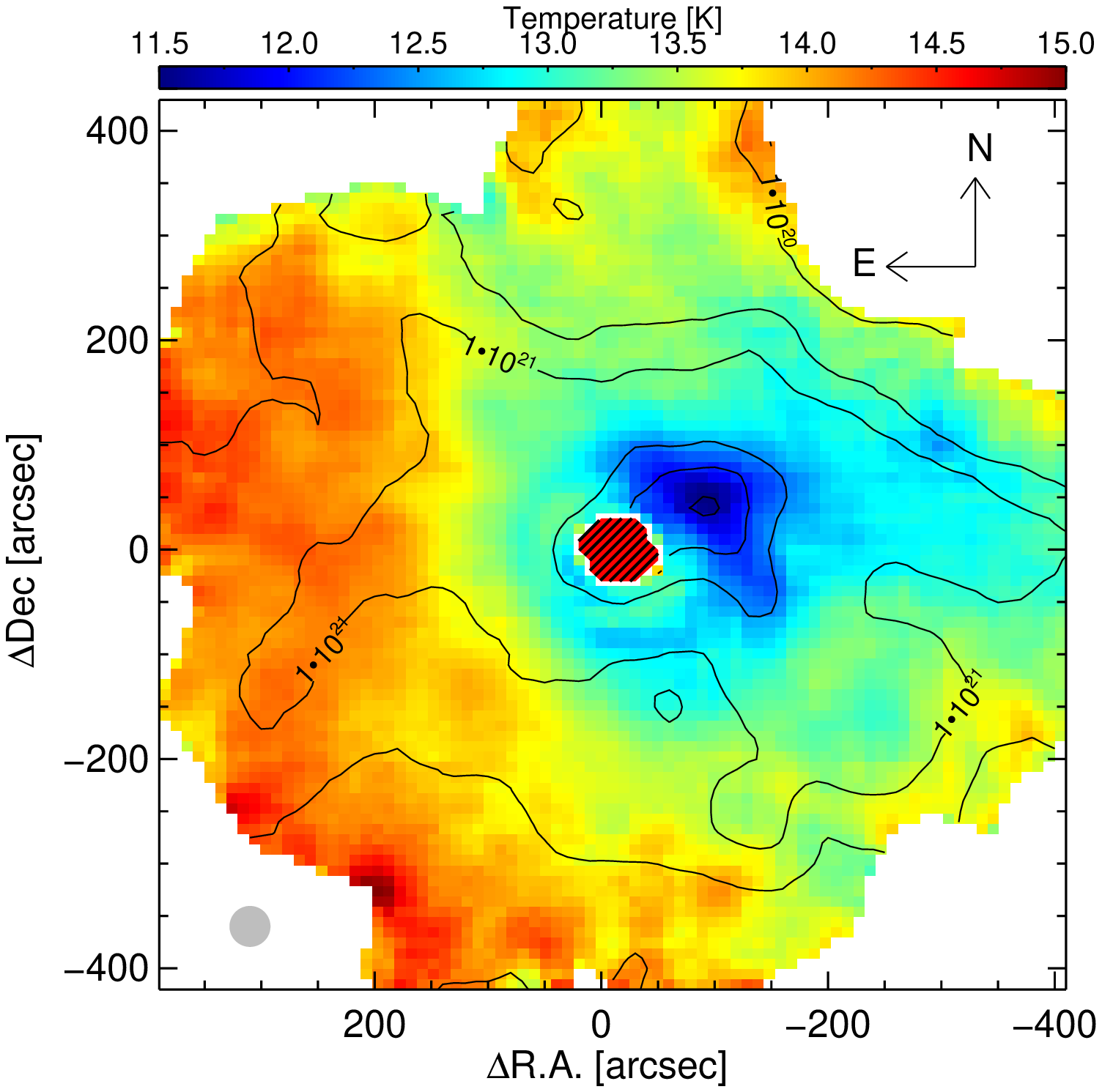}
  \includegraphics[scale=0.5, angle=-90, bb= 88 41 524 500]{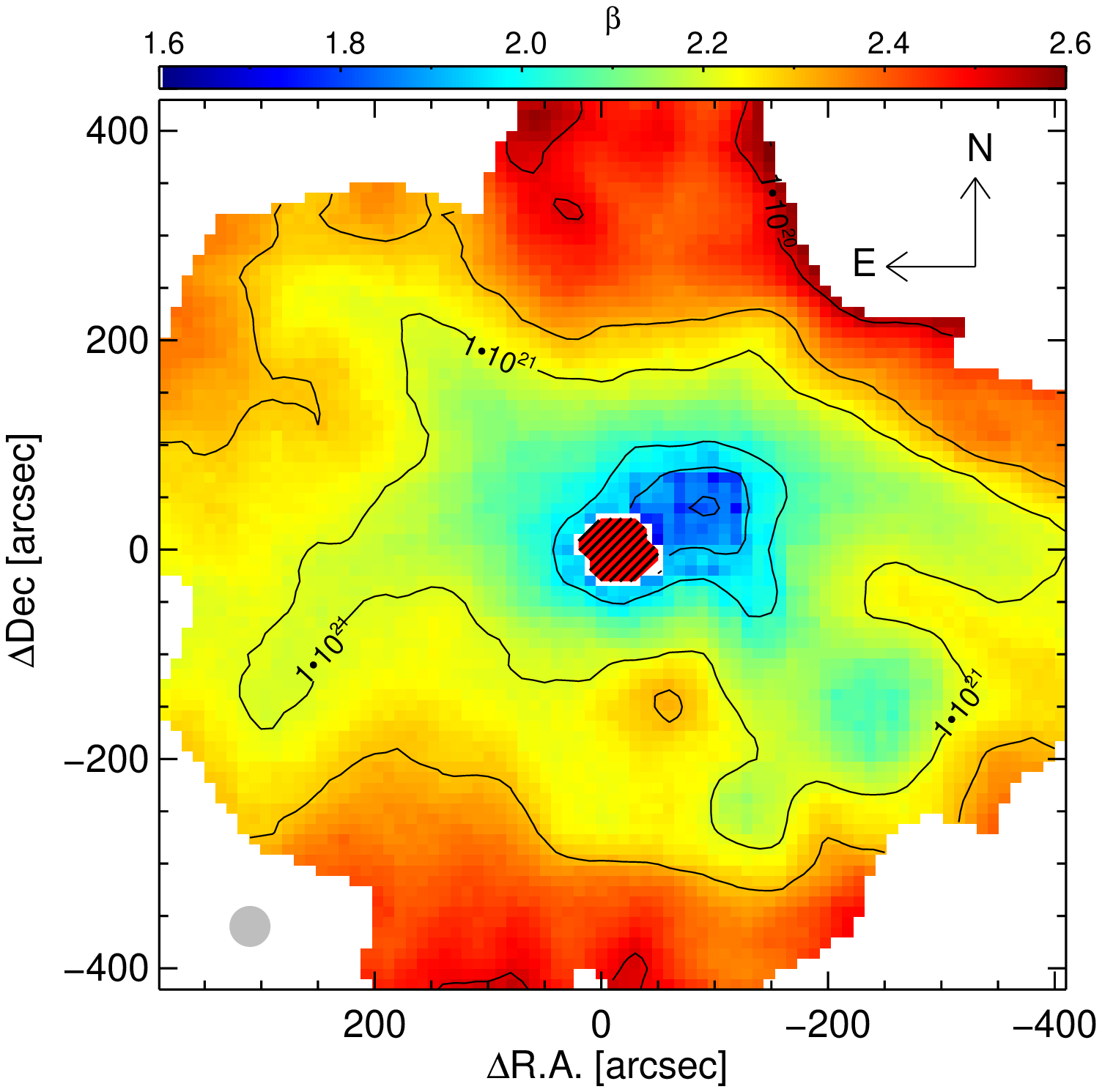}
 \caption{Temperature (left) and $\beta$ (right) map for CB244 derived from the Bayesian
   estimates. The red hashed region in the center of the images corresponds
   to the protostar, which we exclude from our analysis, and the grey
   circle in the lower left corner illustrates the size of the SPIRE
   500~\micron\ beam. Overplotted are contours of constant column
   density, corresponding to $N(H) = 10^{20}, 5 \times 10^{20},
   10^{21}, 5 \times 10^{21}, 10^{22},$ and $2 \times 10^{22}
   {\rm\ cm^{-2}}$. The coolest and most dense region corresponds to
   the prestellar core, with the temperature decreasing toward its
   center.  The $\beta$-map traces the column density map very well, with the
   values of $\beta$ decreasing toward the central, more dense regions.}
 \label{f-cb244_tmap}
\end{figure*}

The temperature and $\beta$ maps of CB244 are shown in Figure
\ref{f-cb244_tmap}, along with contours of constant column density;
all maps were derived from the posterior median values. The
temperature tends to decrease toward the center of the prestellar
core, while the column density tends to increase toward the
center. The $\beta$-map illustrates that the values of $\beta$ trace
the column density values very well, with $\beta$ decreasing toward
the central, more dense regions. The estimated $\beta$ values become
noisy near the central dense region of the core, with more drastic
spatial variations. It is unclear why this is the case, although it
may be related to the breakdown of the assumption of optically thin
isothermal dust that underpins Equation (\ref{eq-modbbody}).

\begin{figure*}
  \includegraphics[scale=0.35,angle=90]{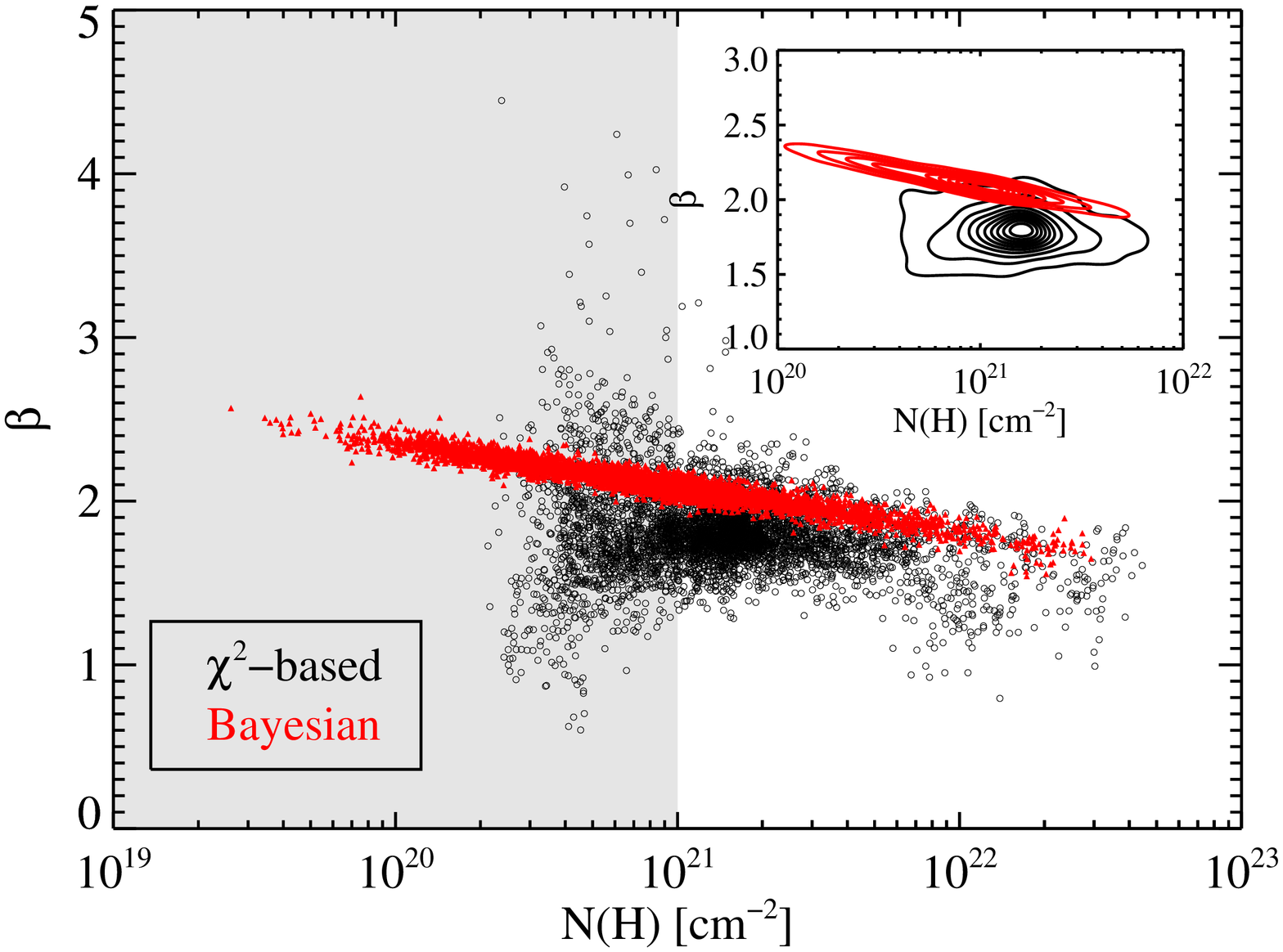}
  \includegraphics[scale=0.35,angle=90]{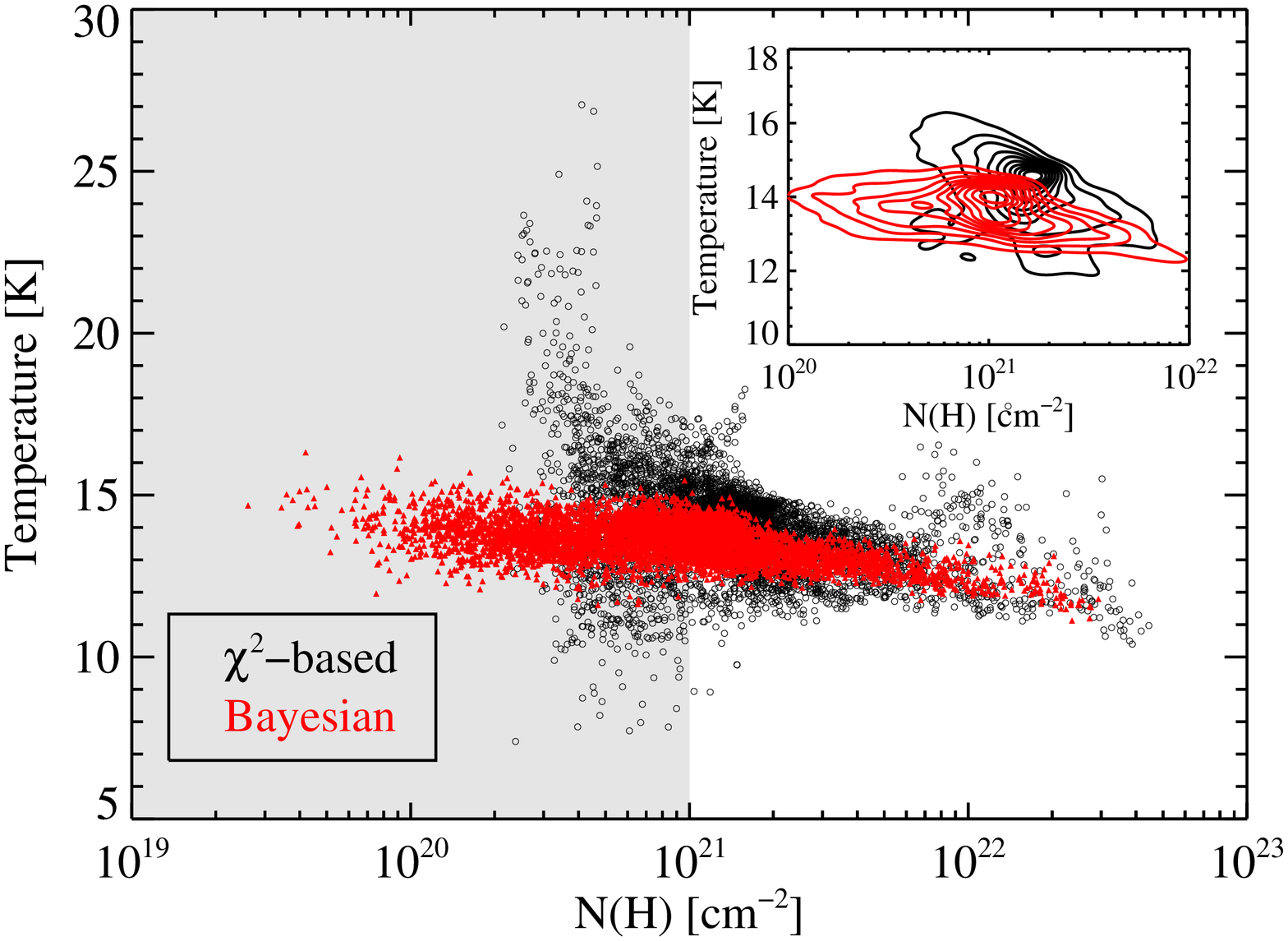}
  \caption{Dependence of $\beta$ (right) and $T$ (left) on column
    density, $N(H)$. There is a tight and highly significant
    anti-correlation between column density and $\beta$, and there is
    a weak anti-correlation between column density and
    temperature. These anti-correlations may be evidence for the
    growth of dust grains in more dense, cooler environments. In both
    cases the trends are not as obvious for the $\chi^2$-based
    estimates because large uncertainties artificially broaden the
    inferred distributions. The estimated values of $\beta$ and $T$
    based on our Hierarchical Bayesian method likely become strongly
    influenced by extrapolation from high $S/N$ data points at $N(H)
    \lesssim 10^{21}\ {\rm cm^{-2}}$ (shaded region), so we caution that the estimated
    trends at low $N(H)$ may not be robust.}
 \label{f-cb244_betaN}
\end{figure*}

Figure \ref{f-cb244_betaN} shows the distribution of the CB244 data
points in the $N(H)$--$\beta$ plane, using both the Bayesian and
$\chi^2$-based estimates. For the Bayesian estimates we show a random
draw from the posterior distribution to more faithfully represent the
intrinsic scatter in $\beta$ at a fixed value of $N(H)$. As expected
from the $\beta$ map shown in Figure \ref{f-cb244_tmap}, our Bayesian
estimates show a tight anti-correlation between column density and
$\beta$. The anti-correlation between $\beta$ and $N(H)$ is also seen
with the $\chi^2$ estimates, but the larger errors in the
$\chi^2$-based estimates add considerable artificial scatter to the
values of $\beta$ at fixed $N(H)$, making the correlation appear less
significant. The correlation can be parameterized as
\begin{equation}
  \beta = (2.18 \pm 0.18) - (0.27 \pm 0.01) \log \left(\frac{N(H)}{10^{21}\ {\rm
        cm^{-2}}}\right) \label{eq-betan_anticorr}.
\end{equation}
The scatter in $\beta$ at fixed $N(H)$ is estimated to have a dispersion
of $\sigma_{\beta|N} = 0.040 \pm 0.003$. This value of
$\sigma_{\beta|N}$ argues against a constant value of $\beta$ at fixed
column density. The anti-correlation between 
$\beta$ and column density is highly significant and is observed even
if we limit our analysis to the highest $S / N$ data.

 The $\beta-N(H)$ anti-correlation obtained from our Bayesian
  method is much tighter than that estimated using the $\chi^2$
  estimates. However, we caution that it is unclear if the small
  scatter in $\beta$ extends over the entire range in column density
  probed. Because $S/N$ is strongly correlated with column density,
  the pixels with low estimated column density also have low
  $S/N$. For the lower $S/N$ pixels, the model for the distribution of
  $\beta,N(H),$ and $T$ becomes more informative, and thus the
  distribution of the low $N(H)$ estimates strongly depends on
  extrapolation from the distribution of the high $N(H)$
  estimates. Therefore, the distribution of $\beta$ at fixed $N(H)$ at
  low $N(H)$ is primarily estimated by extrapolation from the
  distribution of $\beta$ at fixed $N(H)$ at high $N(H)$. Our simple
  Student's t model fixes the standard deviation of the scatter in
  $\beta$ at fixed $N(H)$ to be constant. Because the scatter in
  $\beta$ at fixed $N(H)$ is small for high $N(H)$, where the $S/N$ is
  high, the Student's t-model therefore extrapolates the scatter in
  $\beta$ at fixed $N(H)$ at low $N(H)$ to also be small. 

  The increasing influence of extrapolation on our results can be seen
  more clearly in distribution of $T$ at fixed $N(H)$, where the
  dispersion in estimated $T$ at fixed $N(H)$ increases down to $N(H)
  \sim 10^{21}\ {\rm cm^{-2}}$, an then becomes constant. This
  therefore suggests that the estimated values for pixels $N(H)
  \lesssim 10^{21}\ {\rm cm^{-2}}$ are strongly influenced by
  extrapolation from the pixels with $N(H) \gtrsim 10^{21}\ {\rm
    cm^{-2}}$. If we had 
  used a more flexible model, such as a mixture of Gaussian functions,
  then the high $S/N$ data would not be as informative about the low
  $S/N$ data and the scatter in $\beta$ at fixed $N(H)$ for the low
  $N(H)$ data would be poorly constrained. Because of this, we cannot
  conclude that the tight anti-correlation derived from our
  hierarchical Bayesian approach persists across the entire range of
  column density probed in this analysis, although we do find evidence
  that it is real at $N(H) \gtrsim 10^{21}\ {\rm cm^{-2}}$. Future
  work will incorporate more flexible models for the distribution of
  $\beta,N(H),$ and $T$. However, we also note that the weak positive
  correlation between $\beta$ and $T$ that we observe persists if we
  limit ourselves to only those pixels with $N(H) > 10^{21}\ {\rm
    cm^{-2}}$.

Figure \ref{f-cb244_betaN} also shows the distribution of the CB244
data points in the $N(H)$--$T$ plane, using both the Bayesian and
$\chi^2$-based estimates. As with the $N(H)$--$\beta$ distribution, we
use a random draw from the posterior distribution for the Bayesian
estimates. An anti-correlation between column density and temperature
is apparent, although it is not as tight as the anti-correlation
between $N(H)$ and $\beta$. The anti-correlation between column
density and temperature is also manifested in the temperature map of
CB244 (Fig. \ref{f-cb244_tmap}), where the temperature decreases
toward the central denser regions. However, comparison with the
$\beta$-map in Figure \ref{f-cb244_tmap} shows that values of $\beta
\approx 2.2$ are found both in the warmer region to the east of the
core, and in the cooler region to the west of the core. These results
suggest that density, and not temperature, is the primary driver
behind variations in $\beta$.

Our analysis also enables us to investigate how $\beta$ depends on
temperature at fixed column density. Although we have found that
$\beta$ and $T$ are weakly positively correlated in CB244, this is
obtained by averaging over the distribution of $N(H)$ for
CB244. Therefore, it is not necessarily true that $\beta$ and $T$ are
correlated at fixed $N(H)$. We quantify the relationship between
$\beta$ and $T$ at fixed $N(H)$ by performing a linear regression of
$\beta$ simultaneously on both $\log T$ and $\log N(H)$, which we
derive from our estimated covariance matrix for $\beta,\log T,$ and
$\log N(H)$:
\begin{eqnarray}
  \beta & = & (2.29 \pm 0.18) - (0.29 \pm 0.01) \log \left(\frac{N(H)}{10^{21}\ {\rm
          cm^{-2}}}\right) \nonumber \\
  & - & (0.81 \pm 0.19) \log \left(\frac{T}{10\
      {\rm K}}\right) \label{eq-beta_tn_anticorr}.
\end{eqnarray}
Interestingly, our results imply that for CB244 $\beta$ and $T$ are
anti-correlated at fixed column density. However, when averaging over
the distribution of column densities $\beta$ and $T$ display a weak
positive correlation in CB244. The scatter in $\beta$ at fixed column
density and temperature is $\sigma_{\beta|N,T} = 0.036 \pm 0.002$,
which is only $\approx 10\%$ smaller than the scatter in $\beta$ at
fixed column density. These results confirm that the variations in
$\beta$ are primarily accounted for by variations in $N(H)$, and that
temperature is only a minor secondary driver to variations in $\beta$.

\subsection{Assessing the Impact of Data Systematics}\label{s-cb244_systematics}

As stated above, we limit our analysis to those pixels with $\langle
S/N \rangle > 2$ over the five Herschel bands, with the goal of
minimizing the effect of systematic error in the zero-level of the
maps, which would affect the faint regions the most. However, it is
possible that our results are still affected by systematics regarding
the estimated zero-level. Such systematics may result from
uncertainties in both the astrophysical and instrumental background
levels. In both cases this would result in a spatially-correlated
systematic error in all of the flux values across the map. To assess
the impact of errors in the estimated zero-level, we perform a few
additional checks regarding systematics in the DC-levels. However, we
note that in order to fully realize the impact of data systematics on
the results for a particular source more rigorous simulations should
be performed. As this is beyond the scope of this paper, the tests we
perform here are meant to be illustrative of the impact of data
systematics on the Hierarchical Bayesian results, and how one might
address them in practice. Moreover, we also note that unaccounted
systematics also affect the $\chi^2$-based results in a similar, if
not the same, manner. Thus our result that the Hierarchical Bayesian
method leads to opposite conclusions with respect to the $\chi^2$ estimates regarding the correlations among
SED parameters for CB244 is robust against these systematics.

We first assess the impact of systematic uncertainty in the
zero-levels of the data from ground based bolometers (i.e., SCUBA and
MAMBO). For the ground-based data it is unlikely that the uncertainty
in the zero-level is driven by uncertainty in the background emission,
as the subtraction of atmospheric emission is taken care of in the
measurement procedure. The MAMBO and SCUBA data are chopped at high frequency,
and the restored dual beam maps automatically subtract off all
atmospheric and astrophysical extended background. The information on
the atmospheric emission and extended background are no longer in the data.

In order to assess the robustness of our derived correlations to
systematics with respect to the ground-based data, we redid our
anlysis using only the five {\it Herschel} maps. Using only the \emph{Herschel} data, the
Spearman's rank correlation coefficient for the $\beta$--$T$ relationship was $\rho = 0.37 \pm 0.04$ and
the slope of the $N_H$--$\beta$ anti-correlation was $-0.19 \pm
0.01$. The $\beta$--$T$ correlation is essentially unaffected by the
removal of the ground-based data, while the $N_H$--$\beta$
anti-correlation is reduced in magnitude but still present. Our
results are thus qualitatively robust against the zero–level 
offsets in data from the ground–based bolometers, which frequently
result from the spatial filtering techniques applied during data
reduction \citep[e.g.,][]{kauffmann2008}.

As stated earlier, CB244 is part of a larger sample of Bok
globules chosen to lie in relatively isolated regions. These sources
were chosen in this manner so as to ensure that they have
exceptionally low background emission, and thus uncertainties in the
astrophysical background emission should be minimal. The uncertainty in the
zero-level for the \emph{Herschel} maps should be driven by the
uncertainty in the instrumental background. To assess the impact of mispecifying the zero-levels of the Herschel
data, we refit the CB244 data using only the Herschel bands after
adding Gaussian noise to the logarithm of each of the five estimated DC-levels. The
standard deviation in the log-normal noise was 5\%. This value is
much larger than the formal statistical uncertainty on the DC-level
as estimated according to the procedure described in \S~\ref{s-cb244}
(i.e., on the mean value of the best-fit Gaussian function),
but it should give us insight into how uncertainty in the
DC-level affects the results. Note that each pixel in a map
is assumed to have the same DC-level, and thus the perturbed DC-levels
introduce the same offset to each of the pixels in a given map. 

We did not find any significant
difference to the correlations obtained when fitting the maps with the
perturbed DC levels; however, there is a large
difference in the inferred mean values of $\beta$ and $T$. Using the
perturbed data, our MCMC sampler inferred an average value for $\beta$
and $T$ of $\bar{\beta} = -0.26 \pm 0.32$ and $\bar{T} = 27.3 \pm
0.7$. This value of $\bar{\beta}$ is inconsistent with a value of
$\beta \approx 2$ inferred from our CB244 data and from earlier
studies. Moreover, for the original CB244 data, the best-fit values of
the calibration errors,
$\delta$, were on average 0.7$\sigma$ away from unity, which is
consistent with our prior assumption that the calibration should be
correct on average with an uncertainty of $\approx
15\%$. However, for the perturbed CB244 data the best-fit values of
$\delta$ were on average 1.7$\sigma$ away from unity. The calibration
errors derived from the perturbed data set are inconsistent with our
assumption that they on average should equal unity with a dispersion
of $15\%$, correctly suggesting problems with the perturbed data
set. The likely reason for this is that the error in the zero-level is
partially being absorbed by the estimated values of the calibration
errors, $\delta$. 

As a final test, we fix the calibration errors equal to unity and
refit the data using our Hierarchial Bayesian method. This is
equivalent to assuming that the calibration is correct and that there
is no uncertainty on it. This allows us to test the impact of
overestimating the calibration uncertainty, as we have conservatively
assumed calibration uncertainties that are larger than the official
values recommended by the Herschel Science Center. In addition,
it allows us a more direct comparison with the $\chi^2$ results, as
the $\chi^2$ fits ignore the impact of calibration errors. Fixing
$\delta_j = 1$ only resulted in small differences in the estimated
mean values and correlations of $(\log N_H, T, \beta)$. The derived
value of Spearman's rank correlation for the $\beta$--$T$ relationship was $\rho = 0.28 \pm 0.02$ when
we ignored the calibration errors, compared to $\rho = 0.33 \pm 0.04$
obtained when the calibration errors are included. The
anti-correlation between $N_H$ and $\beta$ is actually stronger when
we ignore the calibration errors, having a value of Corr($\log
N_H,\beta$) = $-0.944 \pm 0.004$, compared to Corr($\log N_H,\beta$) =
$-0.786 \pm 0.040$ obtained when we include the calibration
uncertainties.

When we include the calibration errors the derived mean values are
$\langle \log N_H \rangle = 20.90 \pm 0.12$, $\langle \beta \rangle =
1.92 \pm 0.19$, and $\langle T \rangle = 14.85 \pm 0.42$. However,
when we ignore the calibration errors we find $\langle \log N_H
\rangle = 21.091 \pm 0.006$, $\langle \beta \rangle = 1.797 \pm
0.003$, and $\langle T \rangle = 14.153 \pm 0.016$. Using the median
of the $\chi^2$-based estimates we find $\langle \log N_H \rangle =
21.126 \pm 0.005$, $\langle \beta \rangle = 1.787 \pm 0.002$, and
$\langle T \rangle = 14.265 \pm 0.0179$. Ignoring the calibration
errors produces mean values of the SED parameters and their
uncertainties that are similar to those obtained from the
$\chi^2$-based estimates. In addition, the uncertainties in the means
and correlations of the SED parameters are larger when we include the
calibration errors, as the calibration uncertainty is reflected in the
much larger uncertainties in the SED parameters.

Changing either the DC level or
calibration error results in a constant additive or multiplicative
offset for each map. While these errors can alter the
mean values of the SED parameters, they do not alter their
correlations as the correlations are driven by the spatial
variations of flux values accross the maps, and not by the mean flux
value in each map. More complicated spatially-correlated data
systematics may produce biases in the inferred correlations among SED
parameters. Simulations should be used to assess the impact on the
scientific results when strong spatially correlated errors are thought
to be present in a data set.  

\section{Discussion} \label{discussion} 

The application of our hierarchical Bayesian method to observed fluxes
of CB244 reveals a number of interesting features.  First, we find
that there is only a limited range in \bt\ and \T, with $\beta \in$
(1.8, 2.6), and \T\ $\in$ (11, 16) K.  Second, \bt\ and $T$ are
positively correlated, albeit weakly so, suggesting that the strong
anti-correlation seen in previous work is driven by noise. Further,
the Bayesian fit suggests that \bt\ declines towards the central
region of the starless core, where the temperature decreases and the
column density increases (Fig. \ref{f-cb244_tmap}).  In fact, the
parameterization of \bt\ in terms of \N\ and \T\ in Equation
\ref{eq-beta_tn_anticorr} indicates that \bt\ is more strongly
correlated with \N\ than on \T. 

While we have found a number of interesting features in our analysis
of CB244, their interpretation is more difficult. Strictly speaking,
our derived trends are with respect to the isothermal greybody SED
parameters, which are not necessarily equivalent to the corresponding
physical parameters that they are intended to estimate. Thus, it is
unclear if our derived correlations represent
astrophysically-meaningful correlations, or are instead driven by
systematics involving the data reduction, background subtraction, and
SED model. Such trends may be driven by variations in temperature and
density along the line-of-sight, which are currently not accounted for
in our analysis. For example, the apparent correlation between $\beta$
and $N_H$ cannot be the actual physical correlation as there is both
low and high-density gas along the line of sight. The physical cause
could be a correlation between volume density and $\beta$. In this
sense it is also possible that such trends are at least in part driven
by real astrophysical variations, possibly due to the growth of dust
grains. In high density compressed regions of the ISM, grain sizes may
increase due to dust coagulation, possibly leading to an increase in
\bt.  Compared to the ISM values of \bt\ $\sim$ 2, lower \bt \aplt 1
are found in numerous studies of protoplanetary disks
\citep[e.g.][]{Miyake&Nakagawa93, Mannings&Emerson94, Draine06,
Ricci11}.  The interpretation is that grains in disks are much larger
than in the more diffuse ISM, and that these grains are the seeds of
protoplanets.  The difference in grain sizes between the large scale
ISM and protoplanetary disks suggests that during some epoch of the
star formation process, grains begin to grow. While this is an
intriguing interpretation of our results we stress that we are
currently not in a position to sort out the contributions to our
inferred correlation from the systematics and real astrophysical
variations; future work will address systematic errors resulting from
our assumptions of optically-thin isothermal dust.

In order to accurately map the \T, \bt, or density structure of an
observed region, line-of-sight variations must be taken into
account. For example, \citet{Shettyetal09b} find that when the model
\bt\ is constant but there are temperature variations along the
line-of-sight, the assumption of isothermality produces \bt\ estimates
which are inversely correlated to the fitted
temperatures. Line-of-sight \T\ variations will effect both the
hierarchical Bayesian and $\chi^2$ fits of Equation
(\ref{eq-modbbody}) to the observed fluxes in the same manner. This is
because the Bayesian and $\chi^2$ estimates become equivalent in the
limit of infinite $S/N$.  Because our hierarchical Bayesian model
accounts for the statistical errors, the results obtained from it
should be interpreted as an estimate of what would have been obtained
if there is no measurement error.  It may be that the relationships that we find between
$\beta, T,$ and $N(H)$ are driven at least in part by line-of-sight
variations, making their astrophysical interpretation difficult.

In addition to biases due to line-of-sight variations, our
hierarchical Bayesian results may be biased by the optically-thin
approximation to the SED. We can estimate the magnitude of this bias
using the results from the cross-validation test, described in
\S~\ref{s-cb244_fit}. Because the 100~\micron\ map should be the most
affected by the optically-thin approximation, if the optically-thin
approximation is not valid we might expect the error in the
100~\micron\ data that was omitted from the fit to be systematically
under-- or overestimated. We did not notice any significant offset in
the cross-validation error for either the $\chi^2$ or hierarchical
Bayesian estimates. Moreover, under our assumption of $\kappa_0 =
0.009\ {\rm cm^2 / g}$ with $\nu_0 = 230$ GHz, we estimate the optical
depth at 100~\micron\ in the core to be $\tau \approx 0.05$. Therefore
we do not find any significant evidence that the optically-thin
approximation is having a strong affect on our results.

We can be confident that statistical uncertainties which lead to
spurious and pronounced \Tbeta\ anticorrelations are appropriately
handled in the hierarchical Bayesian method, and thus our inferred
correlations are statistically significant. However, systematic errors
due to mispecification of the SED model, such as line-of-sight
variations, also affect our hierarchical Bayesian results. There may also be
difficulties with the data reduction that can introduce
spatially-correlated systematic errors, such as unidentified
background emission from astrophysical sources, which in turn can bias the
inferred correlations. Therefore at this time we cannot disentagle systematic effects from
real physical effects in our inferred correlations. Nevertheless,
because the hierarchial Bayesian fits rigorously and correctly account
for the statistical errors, we are now in a position to isolate the
effects of systematic errors on the scientific conclusions. A thorough
analysis of possible approaches to account for line-of-sight
variations which does not rely on the optically-thin approximation
will be investigated in a future publication. We will also apply our
method to a large sample of starless cores in order to investigate if
the trends derived here for CB244 extend to a larger sample, providing
a more thorough treatment of data systematics.

\section{Summary} \label{sumsec}

We have developed a hierarchical Bayesian method that rigorously
treats measurement errors for fitting single-temperature greybody
SEDs.  The Bayesian method provides a probability distribution for the
values of temperature \T, spectral index \bt, and column density \N\
in each pixel (or source), conditional on the measured data, as well
as for the distribution of these parameters over an entire map (for a
resolved source) or survey (for multiple unresolved sources).  In
testing the hierarchical Bayesian method on model sources, we
demonstrate that it can accurately recover the true parameters and
correlations, whereas the \chisq\ fit produces an artificial \Tbeta\
anti-correlation due to the degeneracy between \T\ and \bt.

We have applied our hierarchical Bayesian model to {\it Herschel} and
ground-based observations of the Bok globule CB244.  The Bayesian fit
estimates $\beta \in$ (1.8, 2.6), \T\ $\in$ (11, 16) K, which is
significantly more constrained than the \chisq\ estimates.  Further,
we find that \bt\ and \T\ are weakly positively correlated, in direct
opposition to the \chisq\ results.  We have mapped out the spatial
distribution of \T, \bt, and $N_H$, and the correlations between these
properties.  We find that \bt\ decreases from $\sim$ 2.6 where $N_H
\sim 3\times10^{19}\ {\rm cm^{-2}}$, to $\sim$ 1.8 in the densest
region of the starless core, where $N_H \gtrsim 10^{22}$ \cmsq. While
these results may be at least partially driven by systematics
regarding the data reduction and the modeling, our method properly
corrects for the statistical uncertainties, illustrating that the
$\chi^2$ results are significantly affected by noise.

Due to the accuracy of the hierarchical Bayesian method, and its
estimate of a positive correlation between \T\ and \bt, it may be used
to assess any \Tbeta\ anti-correlation found from \chisq\ fits.  Our
analysis demonstrates that hierarchical Bayesian methods can
accurately estimate the dependence between SED parameters, and
therefore may be used to further understand grain evolution in the
ISM.

\section*{Acknowledgements}

We are grateful to Scott Schnee, David Hogg, Karin Sandstrom, Cornelis
Dullemond, Chris Beaumont, Paul Clark, Ralf Klessen, Bruce Draine,
Jonathan Foster, Xiao-Li Meng, Alexander Blocker, and Chris Hayward
for useful discussions regarding dust emission and Bayesian inference.
We are also grateful to an anonymous referee whose suggestions for
additional tests helped improve the paper. This material is based upon
work supported by the National Science Foundation under Grant
No. AST-0908159. BK acknowledges support by NASA through Hubble
Fellowship grants \#HF-01220.01 and \#HF-51243.01 awarded by the Space
Telescope Science Institute, which is operated by the Association of
Universities for Research in Astronomy, Inc., for NASA, under contract
NAS 5-26555, and from the Southern California Center for Galaxy
Evolution, a multi-campus research program funded by the University of
California Office of Research. RS is supported by the German
Bundesministerium f\"ur Bildung und Forschung via the ASTRONET project
STAR FORMAT (grant 05A09VHA), and the SFB 881 ``The Milky Way
System.''

\label{lastpage}
\end{document}